\documentclass[10pt,twocolumn,floatfix,superscriptaddress,amsmath,showkeys,aps,prb]{revtex4-1}
\usepackage{bm}
\usepackage{xcolor}
\usepackage{hyperref}
\usepackage{makeidx}
\usepackage{amsmath}
\usepackage{graphicx}
\usepackage{dcolumn}
\usepackage{amsfonts}
\usepackage{amssymb}
\usepackage{color}
\usepackage{epstopdf}
\usepackage{cancel}
\begin{document}

\title{Excitonic insulators and  Gross-Neveu models}

\author{Nei Lopes}
\email{nlsjunior12@gmail.com (N. Lopes)}
\affiliation{Departamento de F\'{\i}sica Te\'orica, Universidade do Estado do Rio de Janeiro, Rua S\~ao Francisco Xavier 524, Maracan\~a, 20550-013, Rio de Janeiro, RJ, Brazil}
\author{Mucio A. Continentino}
\affiliation{Centro Brasileiro de Pesquisas F\'{\i}sicas, Rua Dr. Xavier Sigaud 150, Urca, 22290-180, Rio de Janeiro , Brazil}
\author{Daniel G. Barci}
\affiliation{Departamento de F\'{\i}sica Te\'orica, Universidade do Estado do Rio de Janeiro, Rua S\~ao Francisco Xavier 524, Maracan\~a, 20550-013, Rio de Janeiro, RJ, Brazil}

\date{\today}

\begin{abstract}

We introduce a generalized Gross-Neveu (GN) model to describe the excitonic instabilities in two different systems: a  small \textit{overlap} semi-metal (SM) and a small \textit{gap} semi-conductor (SMC), both in two (2d) and three-dimensions (3d). We identify the excitonic order parameter (EOP) and obtain the effective potential within the Large $N$ limit approach where the GN model can be exactly solved.   We obtain the excitonic insulator (EI) phase diagrams as a function of temperature, chemical potential, \textit{overlap} between bands and \textit{gaps} of the system. We show that the EI may undergo first- or second-order thermal transitions depending on the regime whereupon this phase is approached. We also investigate the expected thermodynamic signatures for the specific heat above the fine-tuned excitonic quantum critical point (EQCP), in both 2d and 3d, in the SMC regime. We show that the EQCP is a different kind of critical point since although the EOP vanishes at the EQCP, there is always a finite \textit{gap} in the SMC regime. We find that for high temperatures, the specific heat might exhibit a scaling behavior in the form $C_V/T \propto T^{(d-z)/z}$, where $d$ is the dimension of the system and $z$ is the dynamical critical exponent. The very low temperature behavior has a dominant exponential thermally activated term due to the presence of a \textit{gap} that does not vanish at the excitonic transition.\\

\textbf{keywords}: Excitonic insulator, Gross-Neveu model, Large $N$ expansion, phase diagrams.\\

\end{abstract}

\maketitle

\section{Introduction}
\label{sec:Introduction}

In statistical mechanics there are few problems that can be exactly solved. These models are extremely important as they can throw light in relevant aspects of the physics and on the consequences of the different approximations used to deal with similar, but intractable problems. Even when they appear rather unrealistic they still play an essential role in physics.
 
In this work we are particularly interested in a class of models consisting  of $N$ interacting  fermionic fields,  introduced in the 1970s by D. Gross and A. Neveu~\cite{GN}, where the Fermions are coupled by a four-Fermion term satisfying a global $SU(N)$ symmetry~\cite{GN}.  Moreover, the model also exhibits a discrete chiral symmetry.  The Gross-Neveu (GN) model corresponds to a renormalizable version in ($1+1$) dimensions of the Nambu-Jona-Lasinio model~\cite{Nambu} in ($3+1$) dimensions. In this sense, it is widely used by the quantum field theory (QFT) community in the description of quantum cromodynamics (QCD). In the large $N$ limit approach~\cite{Stanley, Wilson, Hooft1, Hooft2}, it is renormalizable in both two (2d) and in its three-dimensional (3d) version. Over the years, the finite temperature properties of the GN model have also been investigated ~\cite{Rudnei, Thies1,Thies2,Thies3, Barducci} as an effective model for both QCD and Fermionic systems in condensed matter physics (CMP). In the latter case, it has applications in areas such as, superconductivity~\cite{Thies2}, polymers~\cite{Caldas} and graphene~\cite{Thies2,Barducci,Rudnei2}.

In a different context, Mott~\cite{Mott} and Knox~\cite{Knox} theorized that in a semi-metal (SM) and in a semi-conductor (SMC), respectively, under certain circumstances, these systems may become unstable to the formation of electron-hole pairs (excitons) giving rise to a new state of matter, the excitonic insulator (EI). The former scenario, described by Mott, is speculated to provide a formal analogy with the Bardeen-Cooper-Schrieffer (BCS) theory of superconductivity~\cite{BCS}, although the physics involved is quite different. Cooper pairs are composed of two electrons and give rise to a supercurrent, while excitons are bounded electron-hole pairs with no net charge. In this sense, the EI state, in general, does not exhibit any special properties concerning the transport of mass or charge~\cite{Moskalenko,Hanamura2}. On the other hand, the scenario outlined by Knox~\cite{Knox}, for the SMC regime, is expected to resemble to the Bose-Einstein condensation (BEC) of a weakly interacting Bose gas. In a SMC, the condition for the electrons in the conduction band and the holes in the valence band to form bound pairs is that the  binding energy of the exciton may exceed the energy gap~\cite{Knox}, such that the ground state becomes unstable against the formation of excitons. Therefore, this instability may appear near to the SM-SMC transition at sufficiently low temperatures~\cite{DesCloizeaux,Keldysh,Jerome,Halperin,Zittartz}. Thus, one can infere that, for solids with small band \textit{overlap} or with \textit{small} energy gaps, there may exist  a new  low temperature  phase of matter, the excitonic insulator~\cite{Franz,Comte,Neuen,nature1,nature2,Zenker,Markiewicz}. 

So, there are two different situations that may lead to the EI phase depending on whether the EI is approached from a SM or a SMC phase. It is worth to emphasize that the nature of the excitonic instability is of great importance, since the EI is a candidate to observe a BCS-BEC crossover in a solid~\cite{Comte,Legget}, which, so far, was only realized in ultracold atomic gases~\cite{Bloch}.

Usually, in the theoretical description of the EI phase~\cite{DesCloizeaux,Keldysh,Jerome, Halperin, Zittartz,Franz,Comte}, the most simple model of two parabolic bands of valence and conduction electrons is considered. The Coulomb interactions between electrons within a band are taken into account through a renormalization of the effective masses of the quasi-particles. In addition, the spin of the electrons is not considered. For simplicity, it is also assumed that the system is isotropic and, in the absence of interactions, has a single valence band with a maximum value in $\vec{k}=0$ and a single conduction band with a minimum value of $\vec{k}= \vec{w}$. Finally, the interacting many-body term between the valence and conduction bands is given by the (partial) charge-density operator.

The search for the EI state in physical systems has been intensive but it is still a topic of debate. Recently, some experimental works~\cite{nature1,nature2,Wakisaka,Eisenstein,Volkov,Sugimoto,disalvo,Kim} have reported measurements on some EI candidate compounds, such as, InAs/GaSb \textit{bilayers}~\cite{nature1,Eisenstein} and Ta$_2$NiSe$_5$~\cite{nature2,Wakisaka,Volkov,Sugimoto,disalvo,Kim}, which may indicate the observation of the EI phase. These measurements revealed a decrease in the valence band below the critical excitonic temperature ($T_c$), which was interpreted as the formation of an additional \textit{gap}, and consequently as the realization of the EI phase~\cite{nature1,nature2,Wakisaka,Eisenstein,Volkov,Sugimoto,Kim}. The relative role of electronic correlations or elastic effects associated with the lattice is still a matter of dispute~\cite{Mazza}.

In order to clarify the nature of the excitonic state,  we introduce in this paper two different models to describe the two distinct regimes where the EI phase is most probable to be found, namely in a small \textit{overlap} SM and in a small \textit{gap} SMC. The models consist of two Dirac bands with correlations described by a quartic interaction as in  GN models. We consider the cases of  both, 2d and 3d, which are solved exactly within the large $N$ limit approach. Besides the usual chiral order parameter (COP), we also introduce an excitonic order parameter (EOP) that characterizes the EI phase. When using the generalized GN model to describe our system, we keep in mind the familiar two-band picture of valence and conduction electrons, assuming that Coulomb interactions between electrons in a given  band  are taken into account by renormalizing these bands~\cite{Jerome,Halperin, Zittartz,Zenker,Markiewicz,Franz,Comte}. For simplicity, the spin of the electrons is not taken into account and we consider the case that the EI phase  emerges from a direct \textit{gap}  system, as for the EI candidate Ta$_2$NiSe$_5$~\cite{nature2,Wakisaka,Volkov,Sugimoto,Kim,Watson}.

We obtain the effective potential in the large $N$ limit~\cite{Stanley, Wilson, Hooft1, Hooft2} for both,  2d and 3d. We show that the EI phase may appear only for a specific range of parameters, depending on the magnitude of the bare \textit{gap} and the relative strength of the chiral and excitonic interactions. In our model, for the SMC case, the EI phase only emerges if the excitonic interaction (binding energy) is larger than the bare and chiral \textit{gaps} of the system, as expected. For the SM case, we introduce a parameter that quantifies the \textit{overlap} between bands. In the latter, the bare and chiral \textit{gaps} are always zero, which means that the \textit{gap} in the EI state appears from a purely excitonic contribution. 

Once we identify the range of parameters that may give rise to the EI phase, we compute the critical exponent $\beta$ for the EOP at zero temperature for 2d as well as 3d case at the SMC regime. We present the EI phase diagrams, in 2d and 3d, as a function of temperature, chemical potential, \textit{overlap} between bands and \textit{gaps} of the system. Our numerical results also show that the system may undergo first- or second-order finite temperature phase transitions, depending on the regime, SMC or SM, whereupon the EI phase is approached. We compare the phase diagrams of both models in 2d and 3d and discuss their main differences depending on the dimension of the system.

We also investigate the expected thermodynamic signatures of the EI transition, obtaining  the specific heat as a function of temperature above the fine-tuned excitonic quantum critical point (EQCP) at the SMC regime for 2d as well as 3d. We show that the EQCP is a different kind of critical point, since although the EOP vanishes at the EQCP, there is always a \textit{gap} (bare and/or chiral) in the SMC regime. In other words, in a small \textit{gap} SMC, when the EI phase disappears, the bare and/or chiral \textit{gaps} remain finite. We obtain, for both 2d and 3d cases, that for high temperatures the specific heat may exhibit a scaling behavior in the form $C_V/T \propto T^{(d-z)/z}$, where $d$ is the dimension of the system and $z$ is the dynamical critical exponent. At low   temperatures the thermodynamic behavior is  exponential thermally activated term due to the presence of a finite \textsl{gap}.

The paper is organized as follows: in Section~\ref{sec:reviewGN} we make a brief review of the two-dimensional GN model pointing out its main aspects from the thermodynamic point of view and emphasizing the features that we are interested in describing the EI state. We also identify the order parameters and the bands structure of our model. In Section~\ref{sec:OP} we present the models to describe the two different regimes expected for the EI phase through the generalized GN model for both 2d and 3d.  We obtain and compare the phase diagrams, for 2d and 3d, for the expected EI phase as a function of temperature, chemical potential, \textit{overlap} between bands and \textit{gaps} of the system, within the Large $N$ limit. In addition, we also investigate the specific heat behavior above the EQCP at the SMC regime for 2d as well as 3d. Finally, in Section~\ref{sec:conclusions} we present our conclusions and summarize the main results.

\section{Brief review of Gross-Neveu models}
\label{sec:reviewGN}

The original version of the  GN model~\cite{GN} describes a   system of $N$-flavored Dirac Fermions in one spatial and one  time dimension,  interacting by means of a scalar-scalar four-Fermion term. The action is given by~\cite{GN,Rudnei,Thies1,Thies2,Barducci,Scherer,Cooper}
\begin{align}
S[\bar{\psi},\psi] = \int d^{2}x \left\{\sum_{j=1}^{N}\left[\bar{\psi}_{j}\left(i\cancel\partial-m\right)\psi_{j}+\frac{G}{2}\left(\bar{\psi}_{j}\psi_{j}\right)^2\right]\right\}
\label{eq: num1}
\end{align}
where $\psi_j$ and $\bar{\psi}_j = \psi^{\dagger}_j\gamma_{0}$ are $N$ independent components Fermion fields $(j=1, . . . , N)$, $\cancel\partial = \gamma^{\mu}\partial_{\mu}$, with $\mu=0,1$,  $m$ is the \textit{mass} and $G$ denotes the coupling constant.  The gamma matrices satisfy $\{\gamma^\mu\,, \gamma^\nu\}=2g^{\mu,\nu}$ with the diagonal metric ${\rm diag}(g)=(-1,1)$ and we choose the following  representation $\gamma_0=\sigma_z$,  $\gamma_1=i\sigma_y$ and $\gamma_0 \gamma_1 = \gamma_5 = \sigma_x$,  where ($\sigma_x, \sigma_y, \sigma_z$) are the usual $2\times 2$ Pauli matrices.

The Lagrangian of Eq.~(\ref{eq: num1}) has a global $SU(N)$ symmetry. In addition, in the massless case, i.e., $m=0$,  the action  is invariant under discrete chiral transformations given by,
\begin{equation}
\psi \rightarrow \gamma_{5} \psi \ \ \ \ \ ; \ \ \ \ \ \bar{\psi} = \psi^{\dagger}\gamma_0 \ \rightarrow \ -\bar{\psi}\gamma_5~.
\label{eq: num2.1}
\end{equation}

However, the \textit{mass} term (bare \textit{gap}) breaks this symmetry since $\bar{\psi}\psi\to -\bar{\psi}\psi$ under this transformation. Thus, chiral symmetry should imply in a gapless spectrum~\cite{GN, Rudnei, Thies1,Thies2,Barducci,Scherer,Cooper}.
It is very well known that the GN model with $m=0$ and  $T=0$ spontaneously develops a \textit{gap} in the spectrum for any finite value of the coupling constant $G$, and at high enough temperature the system closes the \textit{gap}. Therefore, it develops a chiral phase transition at a finite critical temperature. For finite $m$, the chiral symmetry is explicitly  broken and therefore, there is no phase transition. In other words, for $m \neq 0$ there is always a \textit{gap} (bare \textit{gap}) in the system. 

In the same spirit of the original GN model,  it is possible to build a scalar interaction term from the pseudoscalar bilinear $\bar{\psi}\gamma_5\psi$.  This bilinear interaction transforms like the mass term under discrete chiral transformations $\bar{\psi}\gamma_5\psi\to -\bar{\psi}\gamma_5\psi$. Thus, we can write down a generalized chiral GN model as~\cite{GN,Nambu,Flachi}
\begin{equation}
L = \bar{\psi}(i \cancel\partial)\psi+\frac{G_c}{2}(\bar{\psi}\psi)^{2}-\frac{G_e}{2}(\bar{\psi}\gamma_{5}\psi)^{2}
\label{eq:chiralGN}
\end{equation}
where $G_{c,e} > 0$ are two independent coupling constants.   To simplify the notation, we are not explicitly displaying  the index $j$ corresponding to the $N$ Fermion copies. This model has exactly the same symmetries of the original GN model in Eq.~(\ref{eq: num1}) (for the cases with $m = 0$ and $m \neq 0$).

The phase diagram of the generalized GN model in terms of the coupling constants ($G_c,G_e$),  temperature ($T$), and chemical potential ($\mu$) is very rich.  For $G_c> G_e$, the model is completely equivalent to the original GN model, Eq.~(\ref{eq: num1}), displaying a simple phase diagram with a chiral phase transition. The case $G_c=G_e$  is special and it is known in the literature as the Chiral GN model~\cite{GN,Nambu,Flachi}. In this case, the chiral symmetry becomes a continuous symmetry given by the transformation~\cite{Barducci}, 
\begin{equation}
\psi\to e^{i\theta \gamma_5}\psi~.
\end{equation}
Interestingly,  this case (for $N=2$) is tightly related with the Kondo model for a magnetic impurity~\cite{Marino}.

In this paper we are interested in a much less studied case of $G_e>G_c$. We will show that this case is related with the physics of the EI phase~\cite{DesCloizeaux,Keldysh,Jerome,Halperin,Zittartz,Franz,Comte}. Before showing explicit calculations, it is convenient to understand the physics described by the model of Eq.~(\ref{eq:chiralGN}) including the \textit{mass} term ($m \neq 0$). By diagonalizing its quadratic part  in terms of the two components spinor $\psi=(\psi_1,\psi_2)$, we obtain  the dispersion relation for each band given by $\omega (k)=\pm\sqrt{k^2+m^2}$.  Thus, for  $k/m<<1$, we have two parabolic bands separated by a bare gap $2m$, 
\begin{equation}
\omega (k) = \pm \left(m + \frac{k^2}{2m}\right)
\label{eq: num45}
\end{equation}

In the limiting case  $m\to 0$, we recover the linear dispersion,  i.e., $\omega (k)=\pm k$. In this sense, the dispersion relation $\omega (k)=\pm\sqrt{k^2+m^2}$ in the low \textit{momentum} limit, Eq.~(\ref{eq: num45}), describes a SMC system with parabolic valence and conduction bands.

It is interesting to write both scalars appearing in the interaction part,  in terms of the components $\psi_1,\psi_2$.  In the diagonal basis,
\begin{align}
\bar{\psi}\psi&= \psi_{1}^{\dagger}\psi_1- \psi_{2}^{\dagger}\psi_2 
\label{eq:psipsi} \\ 
 \bar{\psi}\gamma_5\psi& =\psi_{1}^{\dagger}\psi_2- \psi_{2}^{\dagger}\psi_1
\label{eq:psigamma5psi}
\end{align}
It is immediate to see from Eq.~(\ref{eq:psipsi}) that $\bar\psi\psi$ is invariant under phase transformation of each band independently.   On the other hand,  from Eq.~(\ref{eq:psigamma5psi}),  $\bar\psi\gamma_5\psi$ is invariant under phase transformation of both bands simultaneously.  This implies  that in the ground state, when the expectation value $\langle\bar\psi\psi\rangle\neq 0$, the charge density of both bands are unbalanced.  However,  charge in each band is conserved. This is the physical content of the chiral symmetry breaking.  On the other hand,  in the case of $\langle\bar\psi\gamma_5\psi\rangle\neq 0$, although the total charge density is conserved, the charge of each band is not conserved , which is analogous to the appearance of a spontaneous hybridization term in CMP systems. This  unbalance of particle-hole condensation between valence and conduction bands characterizes the ground state known as the EI phase.

\section{Order parameters and effective potential}
\label{sec:OP}

In order to compute the thermodynamics properties of the generalized GN model,  Eq.~(\ref{eq:chiralGN}),  with \textit{mass} in (1+1)$d$, we  write the partition function and perform two Hubbard-Stratonovich transformations~\cite{Altland}; one for the coupling $G_c$ with an auxiliary field $\sigma$  and the other for the coupling $G_e$ where we introduce the auxiliary field $\eta$. We find,  
\begin{equation}
Z=\int {\cal D}\bar\psi {\cal D}\psi {\cal D}\sigma{\cal D}\eta~e^{-\int dtdx  {\cal L}(\bar\psi,\psi,\sigma,\eta)} 
\label{eq:Z}
\end{equation}
with
\begin{equation}
{\cal L} = \bar{\psi}(i \cancel\partial-m)\psi-\frac{N}{2g_c}\sigma^2-\frac{N}{2g_e}\eta^2-\sigma\bar{\psi}\psi-i\eta\bar{\psi}\gamma_5\psi
\label{eq: auxiliar_field_SMC_regime}
\end{equation}
where, we have defined the scaled coupling constants $g_{c,e}= N G_{c,e}$.  It is immediate to verify that, integrating out the fields $\sigma$ and $\eta$ we recover the original interaction terms of the generalized chiral GN Lagrangian.

Minimizing the action in Eq. (\ref{eq:Z}) with respect to $\sigma$ and $\eta$, we obtain, 
\begin{align}
\sigma &=-\frac{g_c}{N}\langle \bar{\psi}\psi\rangle 
\label{eq:sigma}  \\  
\eta &=-i\frac{g_e}{N}\langle\bar{\psi}\gamma_5\psi\rangle.
\label{eq:eta}
\end{align}
Comparing Eqs. (\ref{eq:sigma}) and (\ref{eq:eta}) with Eqs.  (\ref{eq:psipsi}) and  (\ref{eq:psigamma5psi}), we  conclude that $\sigma$ is the order parameter of the chiral phase transition, while $\eta$ is the order parameter for the EI phase transition.  
The advantage in using the order parameter fields is that the fermionic integral  in Eq.~(\ref{eq:Z}) is Gaussian and can be done exactly to yield the effective action,
\begin{equation}
S_{\rm eff}=-N{\rm Tr}\ln\left\{i\cancel\partial-(m+\sigma)-i \gamma_5\eta \right\}+\frac{N\sigma^2}{2g_c}+\frac{N\eta^2}{2g_e}.
\label{eq:Seff}
\end{equation}

Note that $S_{\rm eff}$ scales linearly with  $N$. Thus, in the limit $N \rightarrow \infty$ the saddle-point approximation of the partition function becomes exact. If we assume that the order parameters $\sigma,\eta$ are very slowly functions of position and time, we can compute the trace as integrals in frequency and \textit{momentum}. In Euclidean space, performing the trace, we find the effective potential in the large $N$ approximation as follows, 
\begin{equation}
V_{eff}^{N} =\frac{\sigma^2}{2g_c}+\frac{\eta^2}{2g_e}-\int \frac{d^2k}{(2\pi)^{2}} \ln(k^2+\rho^2)+ct
\label{eq_effective_potential_finiteT_SMC_side}
\end{equation}
where $\rho^2 = (m+\sigma)^2+\eta^2$, $k^2=k_0^2+k_1^2$ and $ct$ denotes the \textit{counterterms}, needed for renormalization.
 
The integral in $k$ can be done exactly with an ultraviolet  \textit{cut-off} ($\Lambda$) and,  after renormalization, we obtain
\begin{align}
V_{eff}^{N} =&\frac{\sigma^2}{2g_c}+\frac{\eta^2}{2g_e}+ \notag \\
+&\frac{(m+\sigma)^2+\eta^2}{4\pi}\left(\ln\left[\frac{(m+\sigma)^2+\eta^2}{M^2}\right]-3\right)
\label{effective_potential_zeroT_SMC_side}
\end{align}
where $M^2$ is an arbitrary constant taken as the minimum of the potential~\cite{Nei1}.  

Note from Eq.~(\ref{effective_potential_zeroT_SMC_side}) that the \textit{mass} term $m$  simply shifts the value of $\sigma$, which is consistent with the fact that the \textit{mass} is associated with the bare \textit{gap} of the system, and consequently with the breaking of the chiral symmetry.  Also note that if $m = 0$ and $\eta = 0$ in Eq.~(\ref{effective_potential_zeroT_SMC_side}) we recover the effective potential in the large $N$ limit of the usual GN model~\cite{Rudnei,Barducci}.

The extension  of this formalism to include finite temperatures and chemical potential effects is straightforward using the Matsubara summation technique~\cite{Rudnei,Barducci,Matsubara,Kapusta,Nei2}. We find,
\begin{eqnarray}
V_{eff}^{N}&=&\frac{\sigma^2}{2g_c}+\frac{\eta^2}{2g_e}+ \nonumber
\\
&+&\frac{(m+\sigma)^2+\eta^2}{4\pi}\left(\ln\left[\frac{(m+\sigma)^2+\eta^2}{M^2}\right]-3\right) \nonumber
\\
&-&\frac{T}{\pi}\int_{0}^{\infty}dx \left\{\ln\left(1+e^{-\frac{E-\mu}{T}}\right)+\mu \rightarrow -\mu\right\} 
\label{effective_potential_finiteT_SMC_side}
\end{eqnarray}
where $E^2=x^2+(m+\sigma)^2+\eta^2$.
\begin{figure}[t]\centering
  \includegraphics[width=0.85\columnwidth]{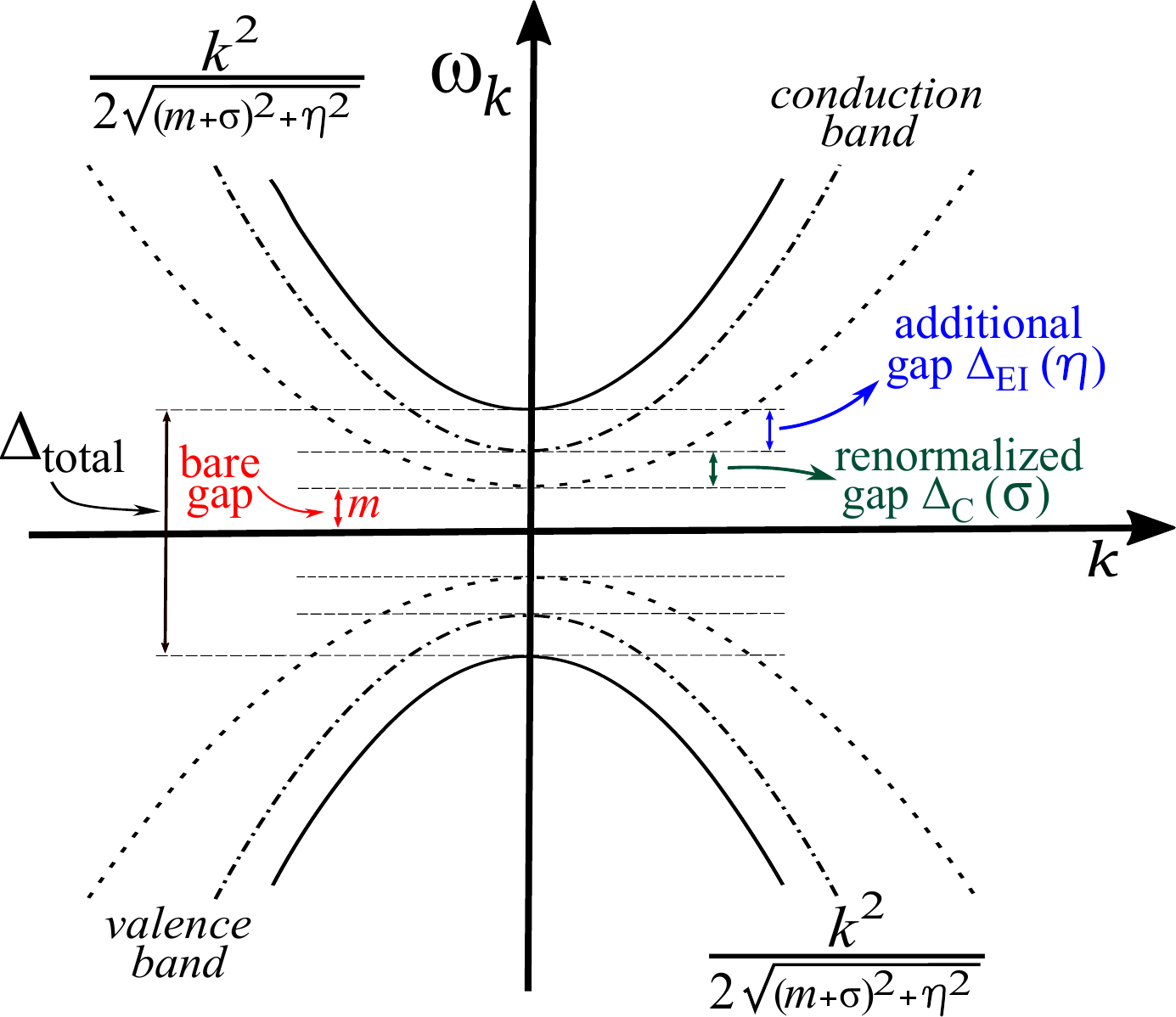}
  \caption{\label{fig:0} (Color online) Schematic for conduction and valence bands from Eq.~(\ref{complete_dispersion}), in the low \textit{momentum} limit. The total \textit{gap} ($\Delta_{\mathrm{total}}$) has three contributions, i.e., $m$, $\sigma$ and $\eta$. The two formers are related to the chiral symmetry breaking in the usual GN model, while the latter is an additional contribution due to the emergence of the EI phase ($\eta \neq 0$) below $T_c$.} 
\end{figure}
This effective potential exactly coincides with the free energy of the system of Eq.~(\ref{eq:chiralGN}) in the $N\to \infty$ limit. The ground state of the model is obtained minimizing Eq.~(\ref{effective_potential_zeroT_SMC_side}) with respect of $\sigma$ and $\eta$.  With these values at hand, we can return to Eq.~(\ref{eq:Seff}) and compute the fermionic spectrum by direct diagonalization.  We get,
\begin{equation}
\omega= \pm\sqrt{k^2+(m+\sigma)^2+\eta^2}.
\label{complete_dispersion}
\end{equation}
Therefore, in the low \textit{momentum} limit,  we find two well defined bands, separated by a total gap $\Delta_{\mathrm{total}}=2\sqrt{(m+\sigma)^2+\eta^2}$, see Fig.~\ref{fig:0}.

Note that, even when the bare dispersion relation is gapless ($m=0$), the ground state becomes gapped, as long as  $\sigma\neq 0$ and/or $\eta\neq 0$. Thus, we identify the phase $\eta\neq 0$ as an EI. The appearance of an additional gap with decreasing temperature, concomitant with a flattening of the bands due to a renormalization of the effective masses is an essential feature of our description of the excitonic state. Furthermore,  this is consistent with the  observation in ARPES measurements~\cite{Wakisaka,nature2}, of an additional gap accompanied by the flattening of the valence (and conduction) band(s) due to the emergence of an EI phase below  $T_c$ in the EI candidate Ta$_2$NiSe$_5$.

\subsection{Phase diagrams of the massive generalized GN model in the semi-conductor regime}

In order to obtain the phase diagrams of the model at zero and finite temperatures, we minimize  Eqs.~(\ref{effective_potential_zeroT_SMC_side}) and~(\ref{effective_potential_finiteT_SMC_side}),  respectively,  with respect to the order parameters $\sigma$ and $\eta$.  In both cases we find a couple of equations that should be solved self-consistently. 
Since we are interested in the EI phase, we consider  $g_e > g_c$.  For instance, for the (1+1)d case, we have considered $g_e = 3.5 > g_c = \pi$.  We have also taken $M=1.0$ in all cases,  which means that all quantities are presented in units of $M$.
\begin{figure}[t]\centering
  \includegraphics[width=\columnwidth]{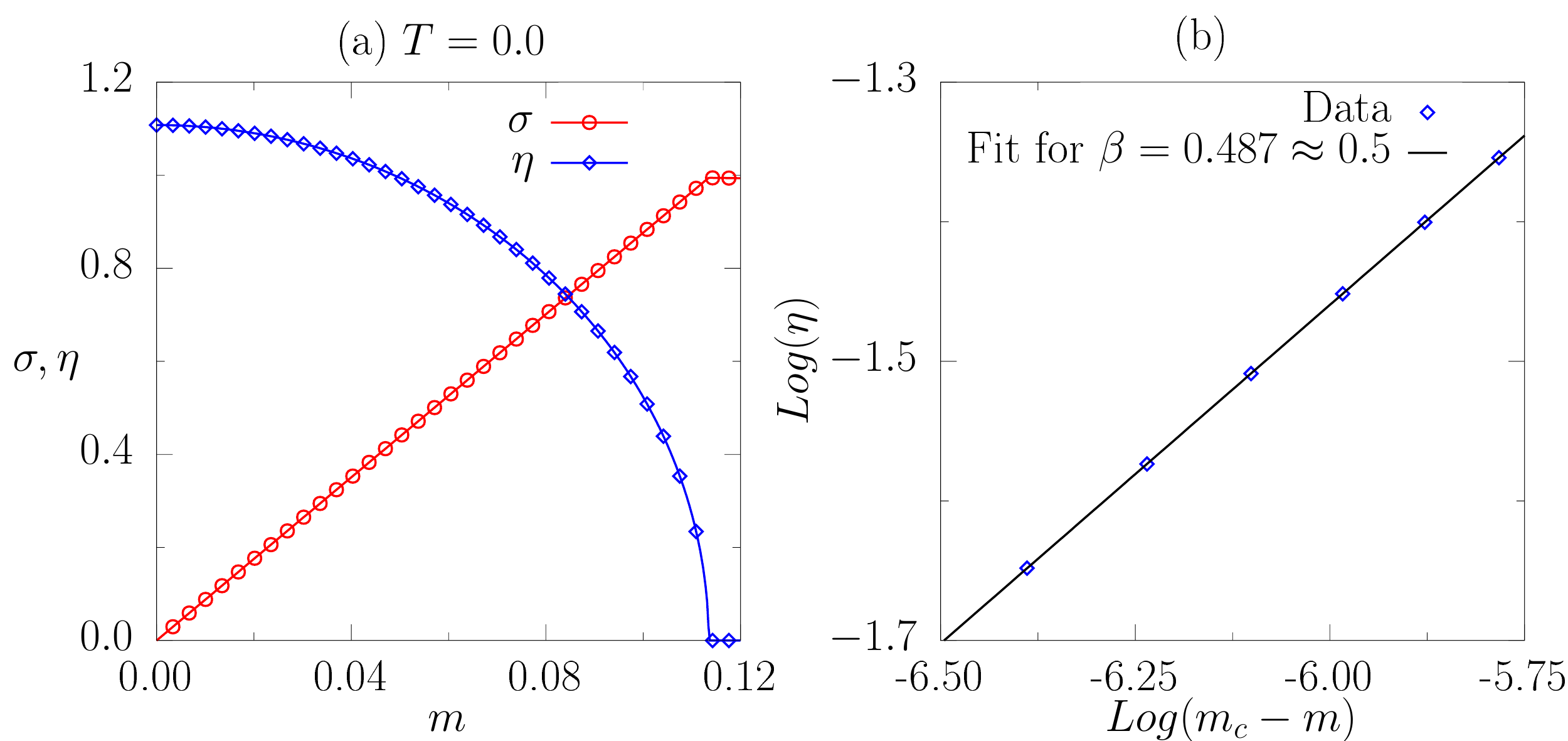}
  \caption{\label{fig:1} (Color online) (a) Order parameters $\sigma$ (red circles) and $\eta$ (blue squares) as a function of \textit{mass} at zero temperature. Note that $\sigma$ is always finite and increases as a function of $m$. On the other hand, $\eta$ decreases and goes to zero continuously as a function of $m$. (b) Using the relation $\eta \propto (m_c-m)^{\beta}$ we obtain $\beta \approx 1/2$ from the fitting curve (solid line) of the Log vs Log plot near the transition point (blue squares data).} 
\end{figure}

In Fig.~\ref{fig:1}~(a) we show the behavior of both order parameters at zero temperature as a function of \textit{mass} (bare \textit{gap}). As expected, $\sigma$ is always finite and increases as a function of $m$.  On the other hand, $\eta$ decreases and goes to zero continuously at a critical mass $m_c = 0.1135$.  The latter behavior is consistent with the fact that  if  the gap of the system increases,  for a fixed value of $g_e$, we expect that above a critical value, the condensation of inter-band particle-hole excitations becomes energetically unfavorable, leading to $\eta=0$. In addition,  we have computed  the critical exponent $\beta$, associated with the vanishing of the EOP, based on the relation $\eta \propto (m_c-m)^{\beta}$.  We have found $\beta \approx 1/2$,  as shown  in Fig.~\ref{fig:1}~(b). This mean-field critical exponent at the quantum excitonic phase transition at $m_c$ for $d_{eff}=1+1$ is a consequence of the large $N$ approximation.

\begin{figure}[t]\centering
  \includegraphics[width=\columnwidth]{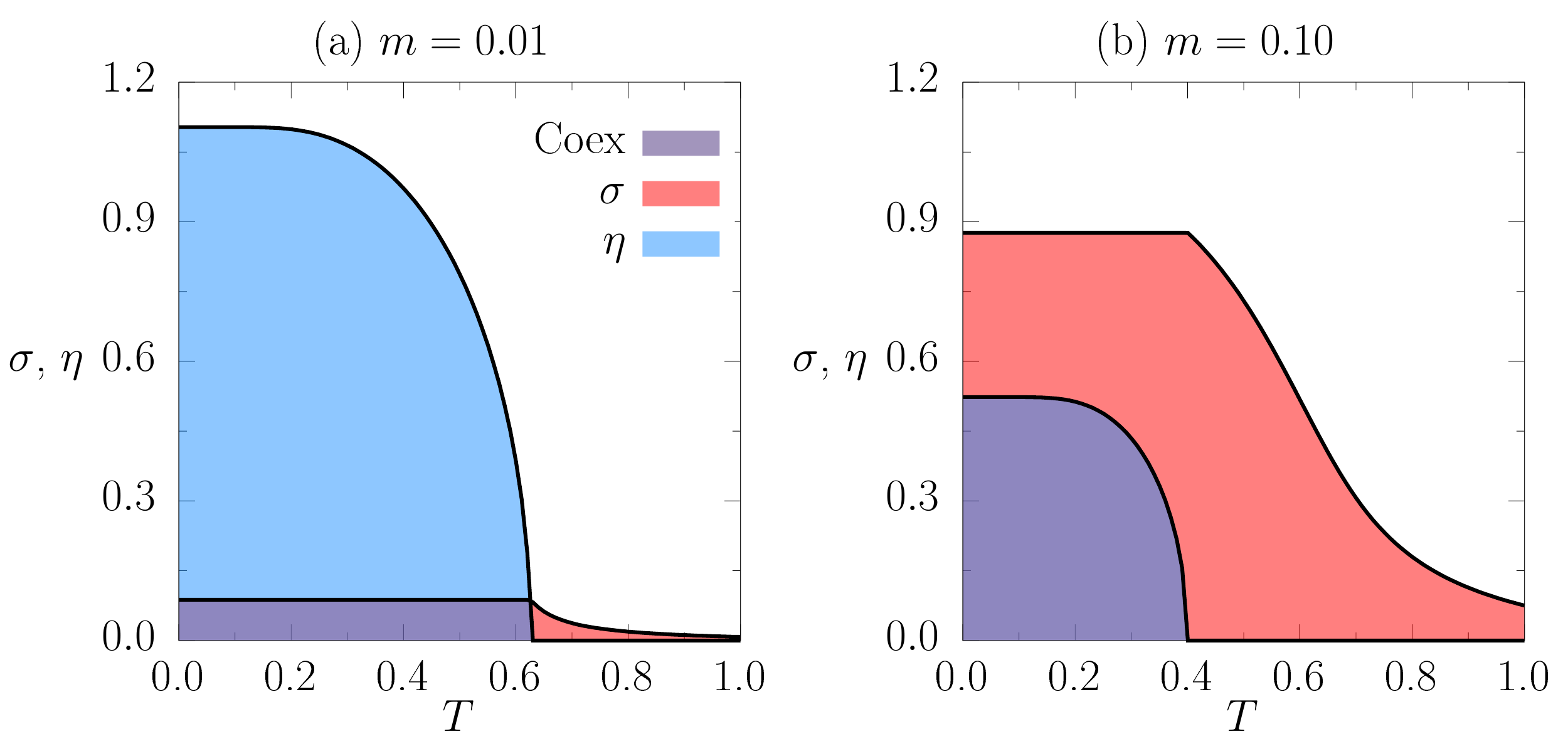}
  \caption{\label{fig:2} (Color online) Order parameters as a function of $T$ for small values of fixed finite mass (bare \textit{gap}). (a) $m = 0.01$, (b) $m = 0.10$. For both plots, $\sigma$ (red) is always finite since $m \neq 0$ breaks the chiral symmetry of the model, while $\eta$ (blue) decreases continuously as a function of $T$. Initially, as a function of $T$, the additional \textit{gap} $2\Delta_{\mathrm{EI}}(\eta)$ shrinks and then, when $\eta = 0$, $\sigma$ starts to decay as a function of $T$, as expected. For large $T$, $\sigma \neq 0$ and it tends asymptotically to the value of the \textit{mass} (bare \textit{gap}), analogously to the usual GN model.} 
\end{figure}

In Fig.~\ref{fig:2}~(a) and~(b), we present the order parameters $\sigma$ and $\eta$ as  functions of temperature for two small values of $m$. In this case,  $\sigma$ is always finite since $m \neq 0$ breaks the chiral symmetry explicitly. In addition,  note that in all the region where $\eta \neq 0$, $\sigma$  is constant. This kind of behavior can be understood from Fig.~\ref{fig:0}. Note that when $\eta \neq 0$ we have the emergence of an additional gap on the system, which is independent from $m$ and $\sigma$, given by $2\Delta_{\mathrm{EI}}(\eta)$. Therefore, as a function of $T$ the additional \textit{gap} $2\Delta_{\mathrm{EI}}$ is first reduced and then, when $\eta = 0$, the renormalized gap ($\sigma$) begins to decrease as a function of $T$, as expected. Finally, at large $T$, the system remains gaped and it tends asymptotically to the value of the bare \textit{gap} (\textit{mass}), analogously to the GN model.  As shown in Fig.~\ref{fig:2}~(b), for larger values of $m$,  the region with finite $\sigma$ increases and that for $\eta$ decreases. Of course, for $m>m_c$, $\eta$ disappears.
\begin{figure}[b]\centering
  \includegraphics[width=0.85\columnwidth]{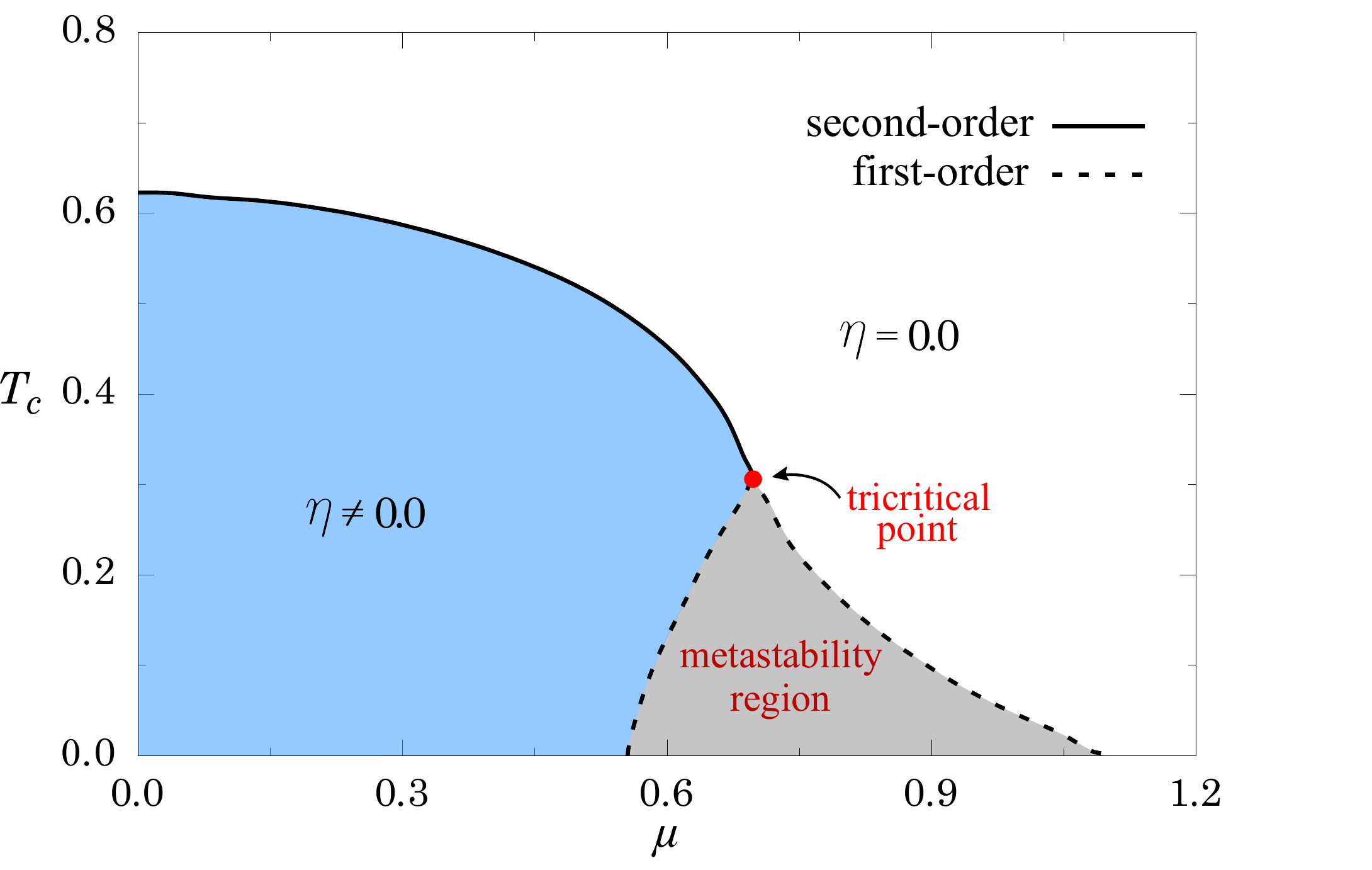}
  \caption{\label{fig:3} (Color online) Critical temperature for $\eta$ as a function of the chemical potential $\mu$ for a fixed $m = 0.01$. Note that this behavior is very similar to the COP $\sigma$ in the usual GN model, where there is a tricritical point (red dot) separating one region with a second-order critical line (continuous line) from a first-order metastability region (dashed).} 
\end{figure}

In Fig.~\ref{fig:3} we show the critical temperature of  the excitonic phase transition  as a function of the chemical potential $\mu$ for a fixed $m = 0.01$.   We observe a behavior quite similar to the chiral phase transition  in the usual GN model,  where there is a tricritical point separating one region with a second-order thermal transition (continuous line) from a first-order one (dashed line)~\cite{Rudnei,Thies1,Thies2,Barducci}.
This phase diagram exhibits metastability in the region between dashed lines for large $\mu$. It is worth to point out that a discontinuous transition for the EI  phase as a function of pressure has been reported for the EI candidate Ta$_2$NiSe$_5$~\cite{nature2}, which may be related to the effects of $\mu$ within our model.

\begin{figure}[t]\centering
  \includegraphics[width=0.85\columnwidth]{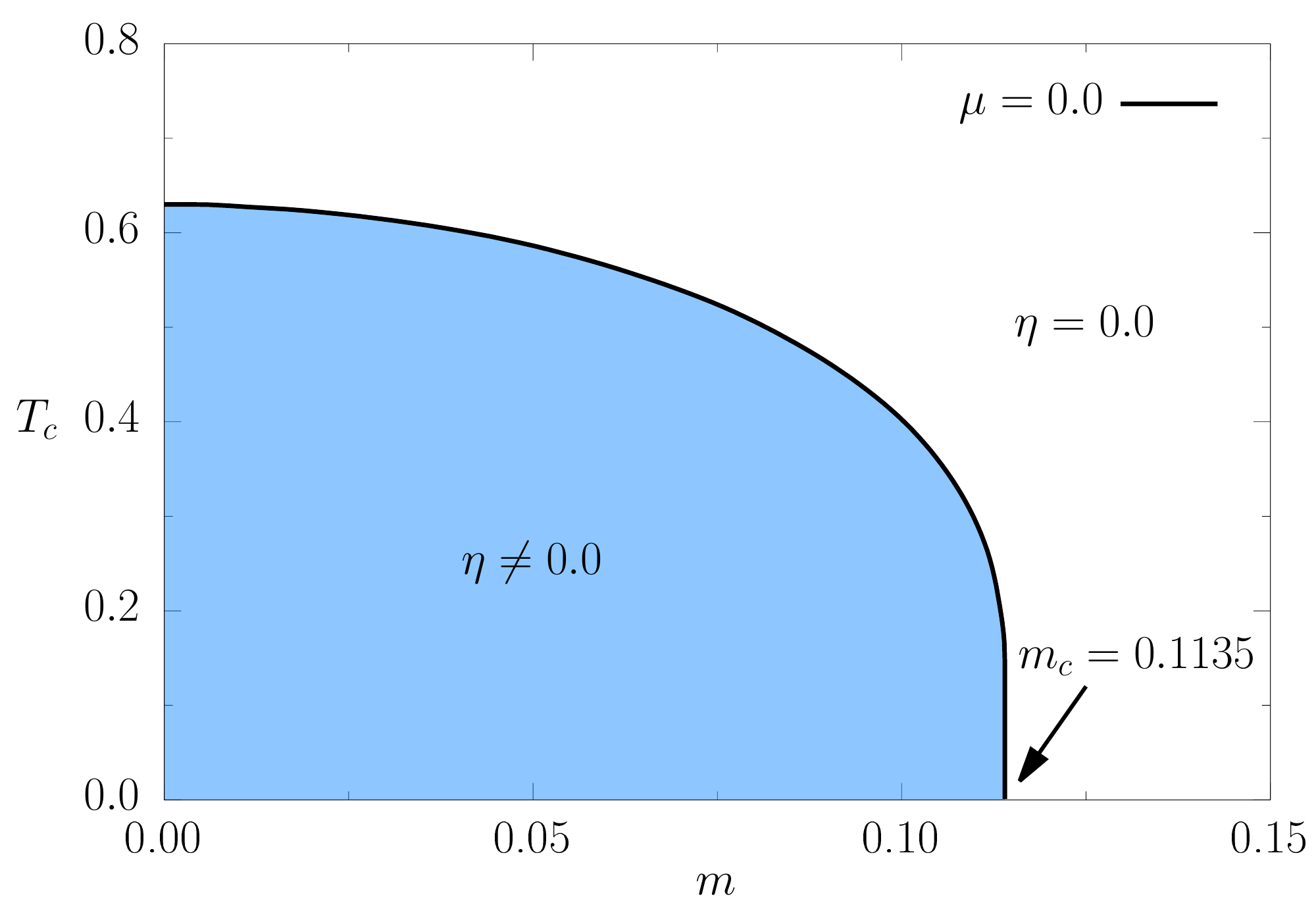}
  \caption{\label{fig:4} (Color online) Critical temperature for $\eta$ as a function of \textit{mass} (bare gap between the bands). There is a critical \textit{mass} ($m_c$) at zero temperature, where the EI phase disappears and this value is independent of the value of $\mu$ since the chemical potential enters in temperature dependent contribution of the effective potential, see Eq.~(\ref{effective_potential_finiteT_SMC_side}). Again, continuous line denote second-order phase transitions.}
\end{figure}

Furthermore,  we have also investigated the critical temperature for $\eta$ as a function of \textit{mass} that is a measure of the bare \textit{gap} between the bands, see Fig.~\ref{fig:4}. One can see that there is a critical \textit{mass} ($m_c$) at zero temperature, where the EI phase disappears and this value is independent of $\mu$. Increasing $\mu$, the critical temperature ($T_c$) shrinks, and by further increasing  $\mu$, the critical line of second-order transitions becomes a first-order one (not shown it this figure).

\subsection{Phase diagrams of the generalized GN model in the semi-metal regime}

So far, we have described the phase transition between a SMC system and an EI state, i.e., SMC/EI. However, when the \textit{gap} between the valence and conducting bands becomes negative, there is a band crossing, where the \textit{gap} closes for definite values of the momenta $\pm k_0$, producing a SM behavior (in the absence of a bare hybridization between the valence and conduction bands). By linearizing the dispersion relation at the crossing points $\pm k_0$,  we can write two independent GN models in the form, 
\begin{equation}
L = \bar{\psi}i \gamma^{\mu}(\partial_{\mu} \mp i K_{\mu})\psi+\frac{G_c}{2}(\bar{\psi}\psi)^{2}-\frac{G_e}{2}(\bar{\psi}\gamma_{5}\psi)^{2}
\label{eq: lagrangeana_SM_side}
\end{equation}
where $K_{\mu} = (0,k_0)$.

\begin{figure}[t]\centering
  \includegraphics[width=\columnwidth]{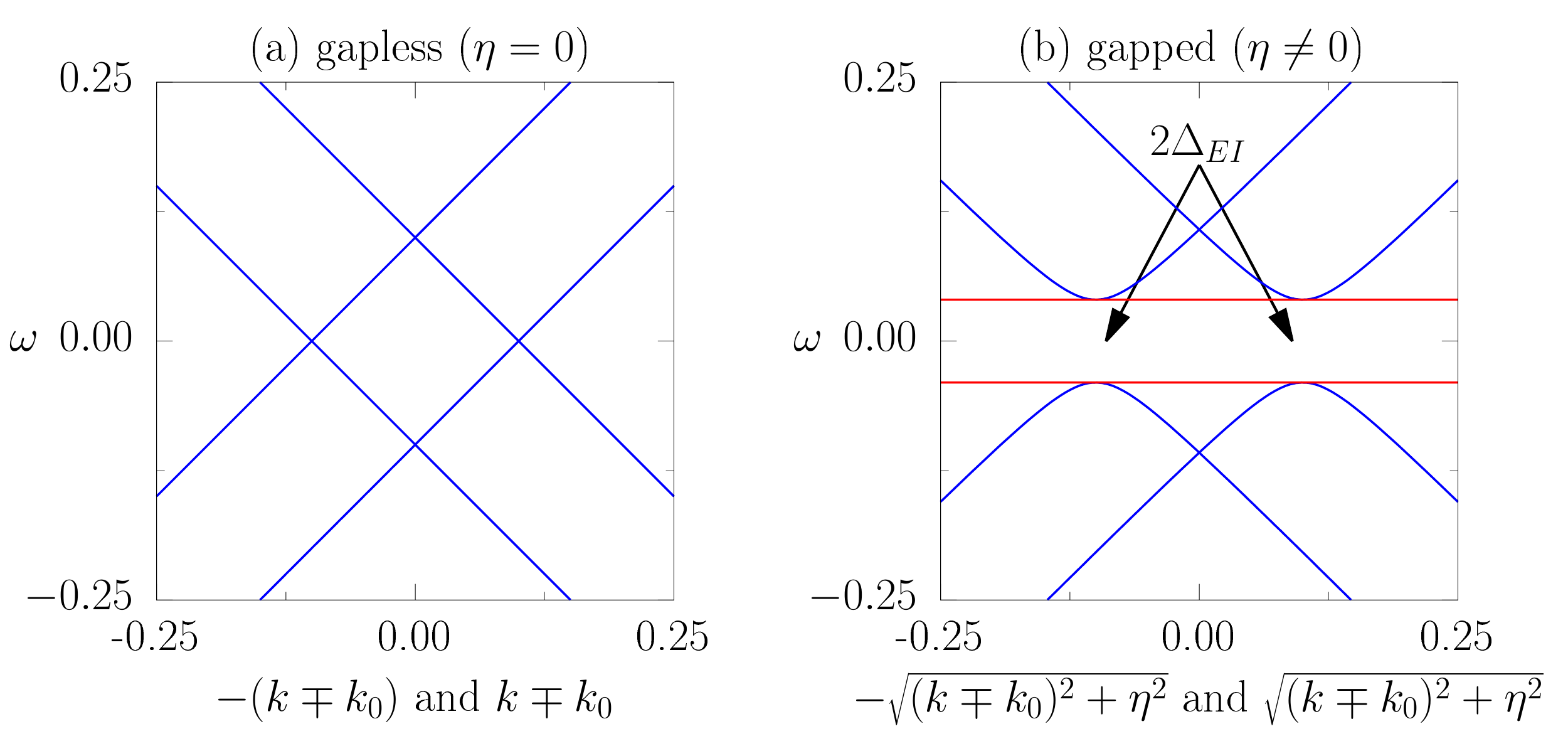}
  \caption{\label{fig:01} (Color online) Dispersion relations for two independent GN models with and without \textit{gap} taking $k_0 = -0.10$. (a) SM-type dispersion relation without \textit{gap}, i.e., an \textit{overlap} between bands. (b) When $\eta \neq 0$ the system exhibits an exclusively excitonic \textit{gap} $2\Delta_{\mathrm{EI}}$.} 
\end{figure}

In Figs.~\ref{fig:01}~(a) and~(b) we depict the dispersion relation of the free term in Eq.~(\ref{eq: lagrangeana_SM_side}).
Note that $k_0$ not only measures the band crossing points, but also the band \textit{overlap}. Indeed $k_0$ is related to the density of carriers in the bands. In 2d spatial dimensions $n=(1/2 \pi^2) k_0^2$. In general, $n \propto k_0^d$ with $d$ the spatial dimension. Then, when $k_0$ is small,  screening is ineffective, the electron-hole attraction is significant  and the excitonic state is favored. For large $k_0$, charge screening is important and this is adverse for the formation of an  excitonic state.  We assume that $k_0$ is small but finite, such that we can ignore interactions between both crossing points $\pm k_0$.  This is a reasonable low energy (long distance) approximation, since these types of interactions involve a $2k_0$ \textit{momentum} transfer. Thus, in this approximation, both models at $\pm k_0$ are decoupled and can be treated as one massless GN model. 

Introducing the same order parameters $\sigma$ and $\eta$, we calculate the effective potential with the same techniques described for the SMC case. Ignoring terms of order $O(k_0^2)$ we find, 
\begin{eqnarray}
V_{eff}^{N}&=&\frac{\sigma^2}{2g_c}+\frac{\eta^2}{2g_e}+\frac{\sigma^2+\eta^2}{4\pi}\left(\ln\left[\frac{\sigma^2+\eta^2}{M^2}\right]-3\right) \nonumber
\\
&-&\frac{k_0}{2}\sqrt{\sigma^2+\eta^2} \nonumber
\\
&-&\frac{T}{\pi}\int_{0}^{\infty}dx \left\{\ln\left(1+e^{-\frac{E-\mu}{T}}\right)+\mu \rightarrow -\mu\right\}
\label{effective_potential_finiteT_SM_side_2d}
\end{eqnarray}
where $k_0 < 0$ and $E^2 = x^2+\sigma^2+\eta^2$.

The phase diagrams are obtained by minimizing Eq.~(\ref{effective_potential_finiteT_SM_side_2d}) with respect to $\sigma$ and $\eta$.  For consistency, we take the same numerical values for $g_{e,c}$ and $M$  used in the discussion for the SMC regime. 

\begin{figure}[t]\centering
  \includegraphics[width=0.85\columnwidth]{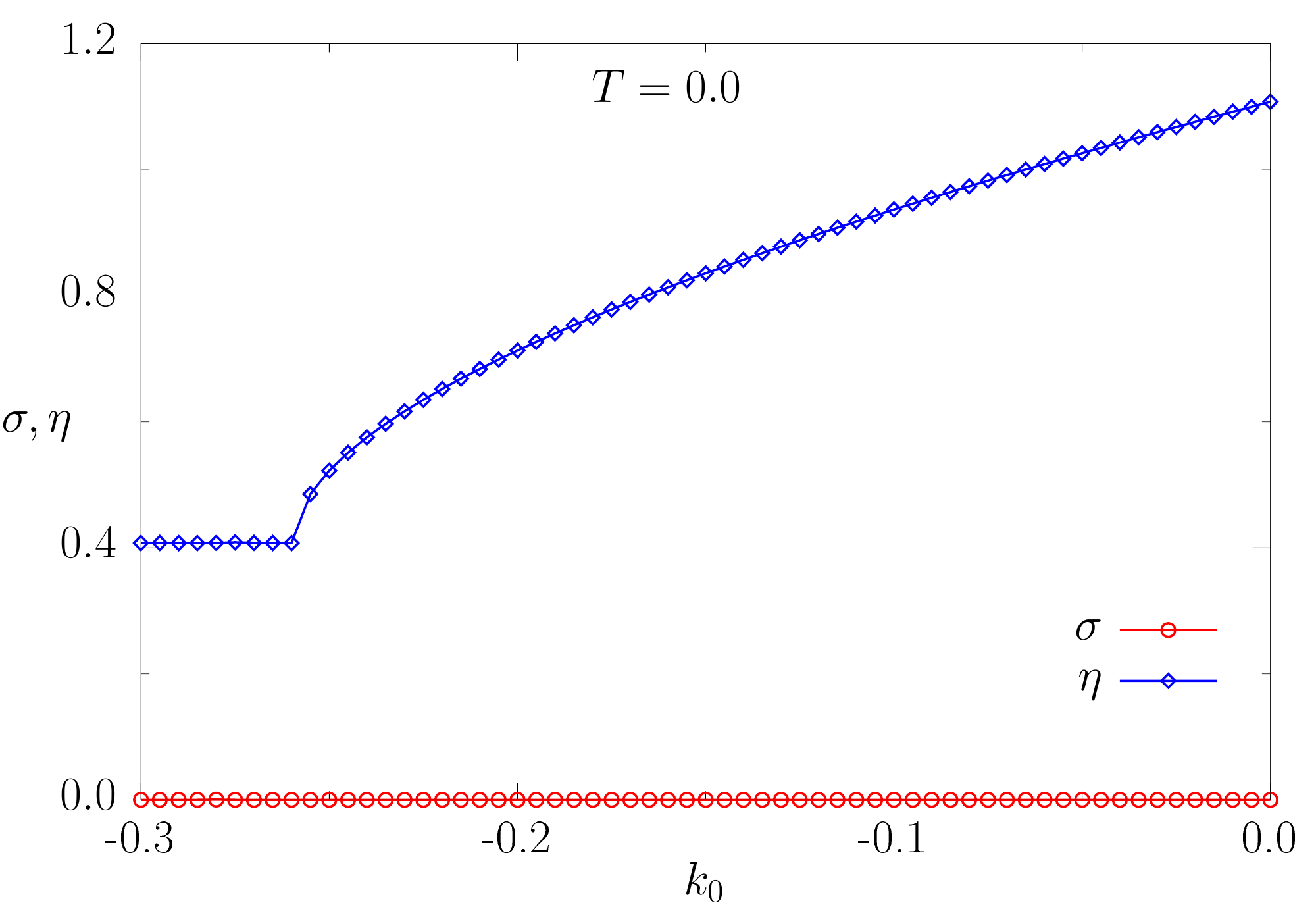}
  \caption{\label{fig:5} (Color online) Order parameters as a function of $k_0$ (overlap) at zero temperature. Note that $\sigma$ (red circles) is always zero, while $\eta$ (blue squares) is always finite in this regime.} 
\end{figure}

Notice that for $m = 0$, the system initially exhibits chiral symmetry due to the \textit{overlap} between bands, as shown Fig.~\ref{fig:01}~(a). The only \textit{gap} that might appear is the excitonic gap in the EI phase. This arises either from $\sigma$ and/or $\eta$, according to the dispersion relation in Eq.~(\ref{complete_dispersion}). In Fig.~\ref{fig:5} we show that in the SM regime at $T=0$,  $\sigma$ is always zero and we have $\eta \neq 0$ for small $k_0 < 0$ that is finite for all $k_0 $ investigated. This means that the \textit{gap} that emerges in the SM regime has an  exclusively excitonic character, see Fig.~\ref{fig:01}~(b).
\begin{figure}[b]\centering
  \includegraphics[width=0.85\columnwidth]{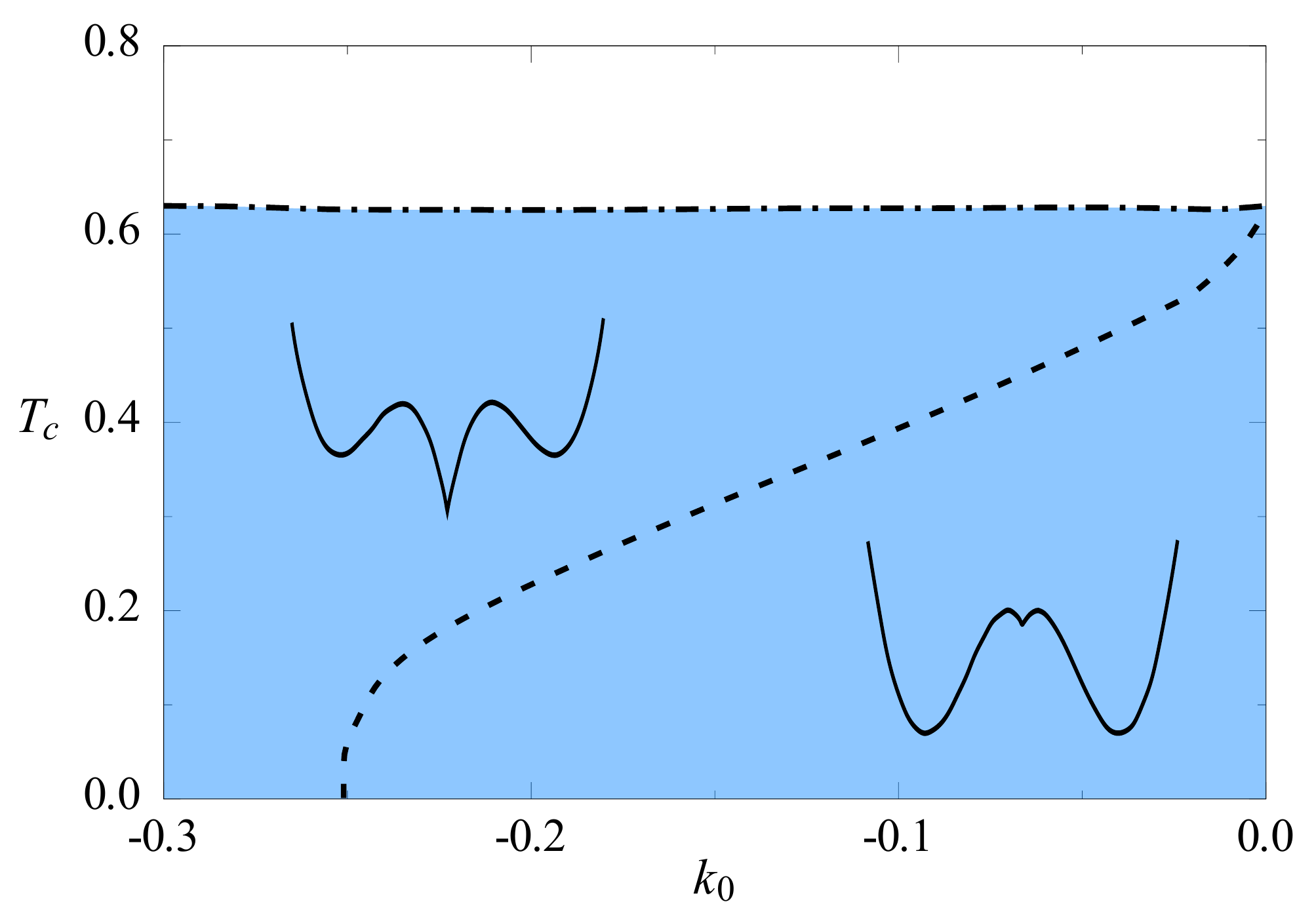}
  \caption{\label{fig:6} (Color online) Critical temperature for $\eta$ as a function of $k_0$ (\textit{overlap} between bands). Note that $\eta \neq 0$ and $\eta = 0$ change stability since the minimum at the origin of the effective potential is always present for $k_0 \neq 0$.  The horizontal dot dashed line is the spinodal line where the metastable excitonic states emerge.} 
\end{figure}

As can be seen in Eq.~(\ref{effective_potential_finiteT_SM_side_2d}),  $k_0$ couples with $\left|\eta\right|$, since $\sigma = 0$ for the SM regime (see Fig.~{\ref{fig:5}}). As a consequence, for $k_0 \neq 0$, the effective potential always displays a minimum at the origin, which competes with the minimum at finite $\eta$. In other words, all SM regime is a metastable one where there is a competition between $g_e$ and $k_0$. The former tends to give rise the EI phase, while the latter acts in detrimental to the EI phase, as expected for the \textit{overlap} between bands.

In Fig.~\ref{fig:6} we show the critical temperature for $\eta$ as a function of $k_0$. In this regime, the critical line is a first-order one (dashed line) for all values of $k_0 < 0$ investigated, where $\eta \neq 0$ and $\eta = 0$ change stability since the minimum at the origin of the effective potential will be always present for $k_0 \neq 0$.

\begin{figure}[t]\centering
  \includegraphics[width=0.85\columnwidth]{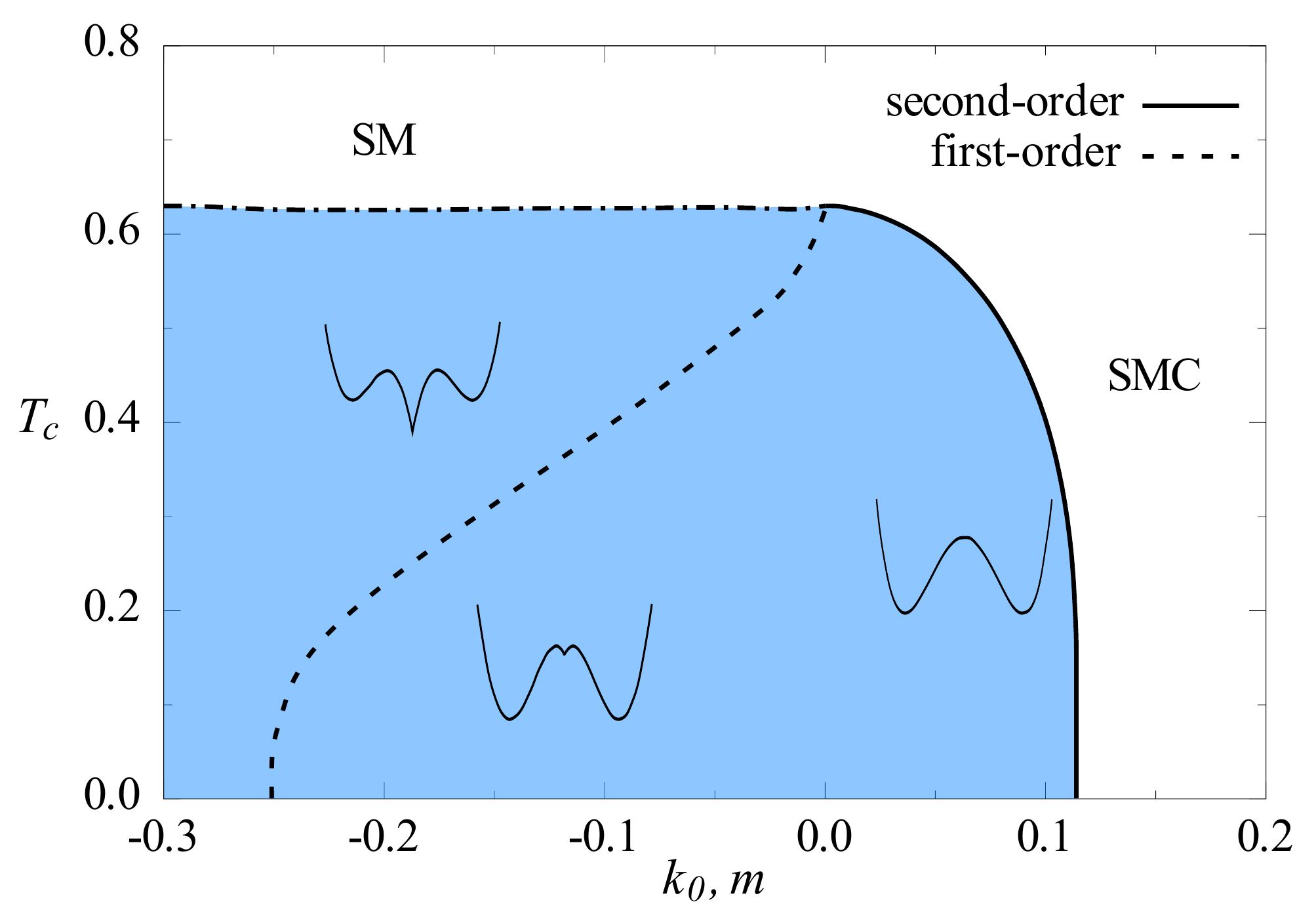}
  \caption{\label{fig:7} (Color online) EI phase diagram combining both 2d models for SMC and SM regimes. In the x-axis we have $k_0$ (negative) for the SM regime, associated with the \textit{overlap} between bands, and $m$ (positive) for the SMC regime, which is related to the bare \textit{gap} of the system.  The dashed curve denotes first-order transitions, while continuous lines describe second-order ones. The horizontal dot dashed line is the spinodal line where the metastable excitonic states emerge.} 
\end{figure}

Combining the phase diagrams of Fig.~\ref{fig:4}, for $\mu = 0$, and Fig.~\ref{fig:6} one can obtain the complete  EI phase diagram within our two-dimensional models, see Fig.~\ref{fig:7}. Note that in the $x$-axis we have $k_0$ for the SM regime, associated with the \textit{overlap} between bands, and $m$ for the SMC regime, which is related to the bare \textit{gap} of the system for fixed $g_{c,e}$. 
Also note that our model can capture the step-like expected shape of the EI phase diagram, which differentiates the two regimes expected for the EI phase, in a qualitative agreement with different approaches~\cite{Franz,Hulsen} although a first-order transition region appears at the SM regime within our model.
From Fig.~\ref{fig:7} one can see the existence of an excitonic quantum critical point (EQCP), in the SMC regime of the EI phase diagram, at a critical mass ($m_c = 0.1135$), see Fig.~\ref{fig:4}. 
\begin{figure}[b]\centering
  \includegraphics[width=0.85\columnwidth]{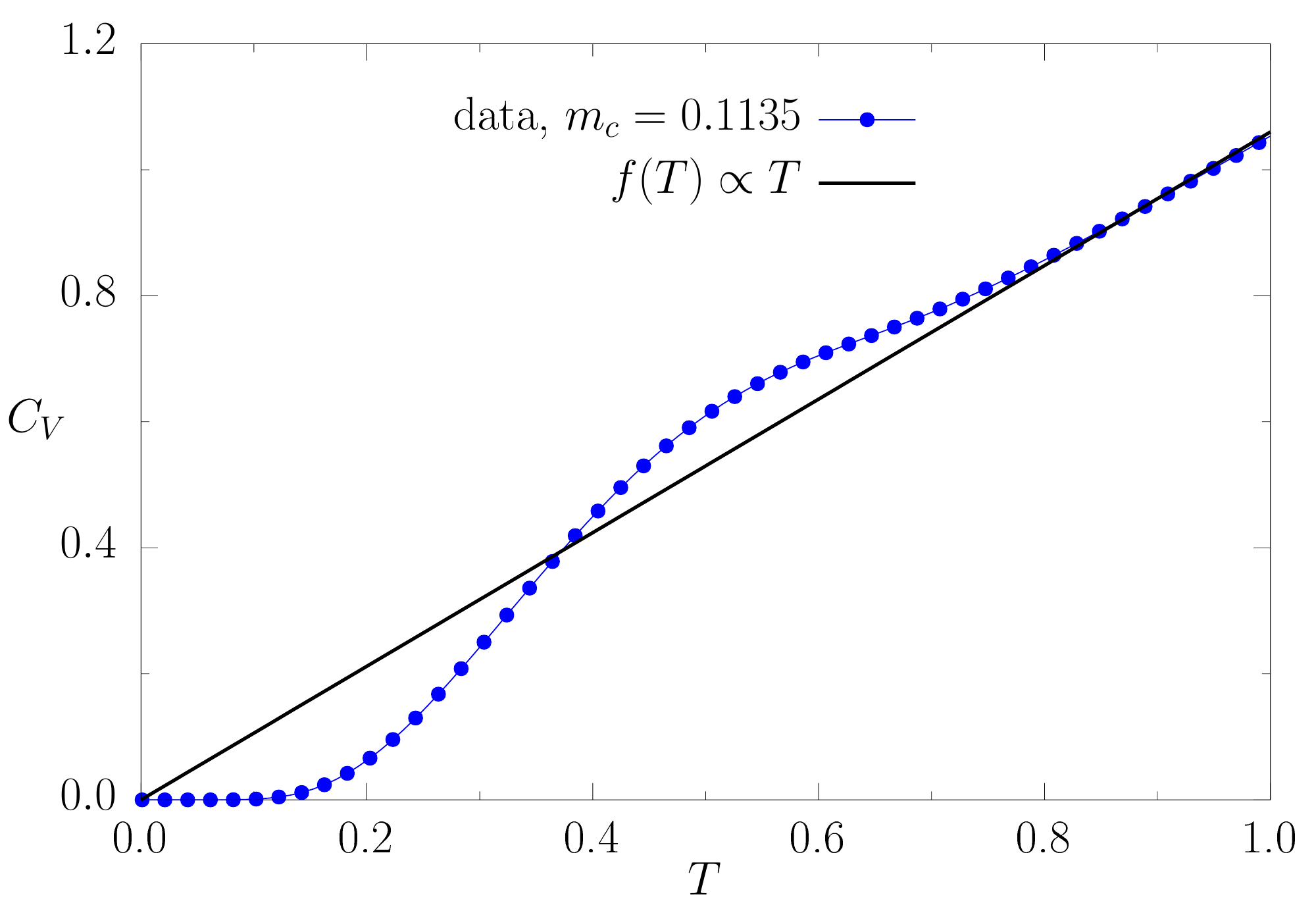}
  \caption{\label{fig:specific_heat_2d} (Color online) Specific heat, at constant volume, as a function of $T$ above the EQCP, i.e., at the fine-tuned value of $m_c=0.1135$. One can see that for large $T$, $C_V$ exhibits a linear behavior,  consisted with the expected behavior at a QCP, $C_V/T \propto T^{(d-z)/z}$ with $d=z=1$. However, at low temperatures the specific heat is thermally activated due to the presence of the \textit{gaps} ($\sigma \neq 0$ and $m \neq 0$), deviating from the linear behavior.}
\end{figure}

From the experimental point of view it is interesting to obtain the thermodynamic signatures above this fine-tuned point for finite temperatures. Thus, in Fig.~\ref{fig:specific_heat_2d} we investigate the specific heat at constant volume, i.e., $C_V = -T \left(\frac{\partial^2 V_{eff}}{\partial T^2}\right)_V$, as a function of $T$ at $m=m_c$.
We have found  a linear behavior at high temperatures, consisted with the expected scaling behavior at a QCP, $C_V/T \propto T^{(d-z)/z}$ with $d=z=1$, where $d$ is the dimension of the system and $z$ is the dynamical critical exponent~\cite{Mucio_book}.  However, at low temperatures, an additional contribution to the specific heat is thermally activated due to the presence of \textit{gaps} ($\sigma \neq 0$ and/or $m \neq 0$). Therefore, the low temperature regime has a dominant  exponential thermally activated term in addition to  the power law contribution due to quantum critical behavior. Then, one can conclude that we are dealing with a different kind of QCP, since although the EOP vanishes at $m_c$ for $T = 0$, there are always \textit{gaps} that break the chiral symmetry  at the SMC regime. That is, while $\eta$  exhibits a critical behavior, $\sigma$ contribute with non-critical fluctuations for a fixed bare \textit{gap} ($m$) at the SMC regime of the EI phase.

\subsection{The Gross-Neveu model in (2+1)d and the excitonic insulator phase}

In this section we compute the phase diagram of the generalized GN model in $(2+1)d$. There are essentially two differences in the definition of the model and in the computation of the effective potential.   On the one hand,  we can no longer  define a chiral operator $\gamma_5 = \gamma_0\gamma_1$ in euclidean 3d.  In that sense,  we need to consider the term of excitonic interactions,  given by $\frac{G_e}{2}(\psi\gamma_1\bar{\psi})^2$,  where $\gamma_1=\sigma_x$ to take into account the same physics given in Eqs.~(\ref{eq:psipsi}) and (\ref{eq:psigamma5psi}) for the COP and the EOP in the diagonalized basis, respectively. On the other hand,  all the integrals over euclidean \textit{momentum} are now given by $\int \frac{d^3 k}{(2\pi)^3} = \frac{4\pi}{(2\pi)^3} \int dk \ k^2$. 

The  solution of the massive GN model  in (2+1)d in the large $N$ approximation is known~\cite{Scherer,Cooper}. Moreover, using the same techniques described before for the $(1+1)d$ version of the model, we can compute initially the effective potential in terms of the order parameters $\sigma$ and $\eta$ for the SMC regime. We get, 
\begin{eqnarray}
V_{eff}^{N}&=&\frac{\left[(m+\sigma)^2+\eta^2\right]^{3/2}}{6 \pi}+\frac{1}{2}\left(\frac{\sigma^2}{g_c}-\frac{(m+\sigma)^2}{g_\Lambda}\right) \nonumber
\\
&+&\frac{1}{2}\left(\frac{1}{g_e}-\frac{1}{g_\Lambda}\right)\eta^2 \nonumber
\\
&-&\frac{T}{\pi}\int_{0}^{\infty}dx \ x\left\{\ln\left(1+e^{-\frac{E-\mu}{T}}\right)+\mu \rightarrow -\mu\right\}
\label{effective_potential_finiteT_SMC_side_3d}
\end{eqnarray}
where $g_\Lambda = 3\pi^2/(2\Lambda)$ and $E^2 = x^2+(m+\sigma)^2+\eta^2$.
In this case, the effective potential presents a natural ultraviolet \textit{cut-off} $(\Lambda)$,  which is encoded in $g_\Lambda$.  From the term proportional to $\eta^2$ in Eq. (\ref{effective_potential_finiteT_SMC_side_3d}),  it is clear that,  to have an excitonic condensate at zero temperature,  we need to fix  $g_e > g_\Lambda$. Without loss of generality, we investigate the phase diagrams by fixing $g_e > g_\Lambda > g_c$.
\begin{figure}[t]\centering
  \includegraphics[width=\columnwidth]{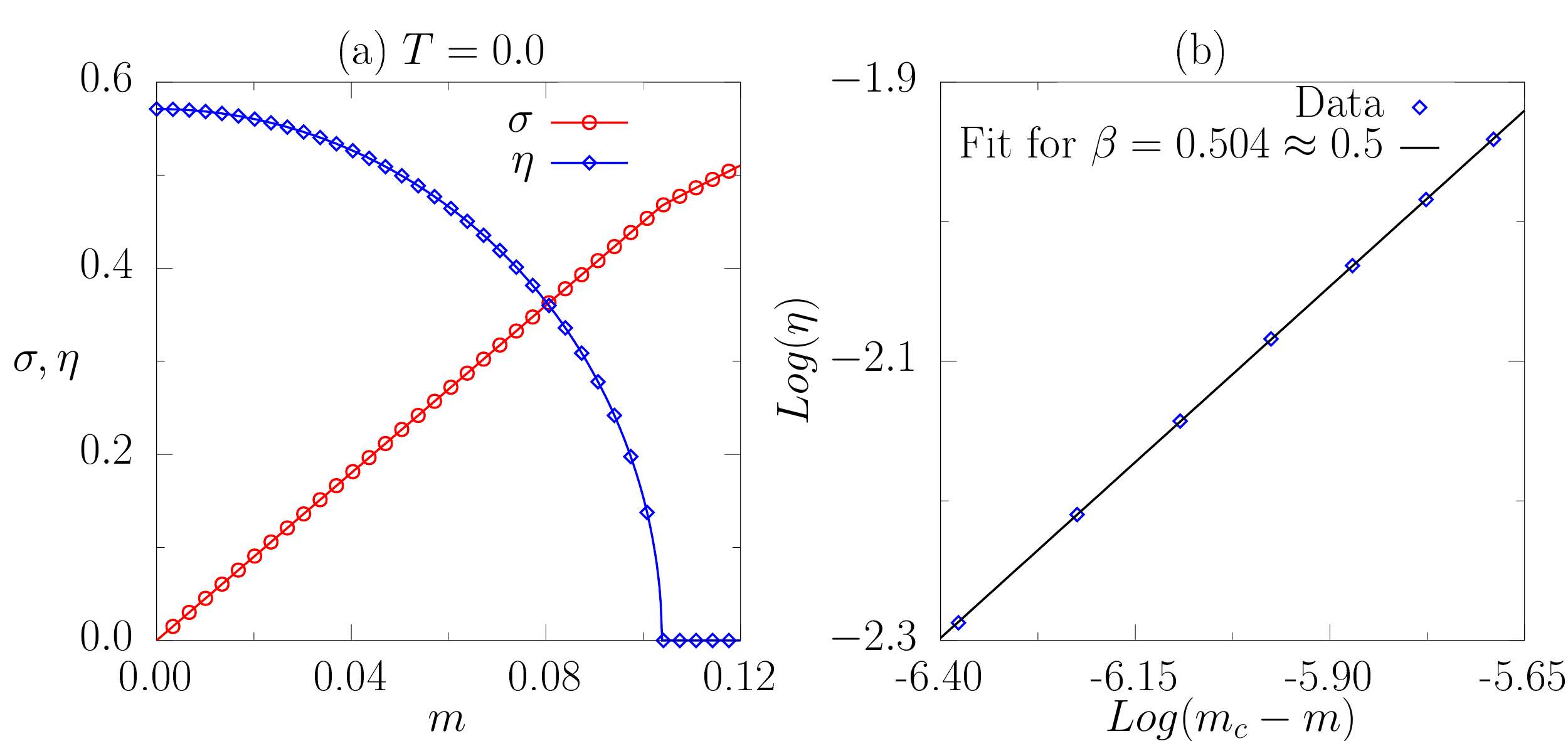}
  \caption{\label{fig:8} (Color online) (a) Order parameters as a function of mass (bare \textit{gap}) at zero temperature for (2+1)d. One can see a very similar behavior for $\sigma$ and $\eta$ when compared to the (1+1)d case (Fig.~\ref{fig:1}~(a)). (b) Analogously to the (1+1)d case, we can obtain the numerical value of the critical exponent $\beta = 0.504 \approx 0.5$ for $\eta$, see Fig.~\ref{fig:1}~(b).} 
\end{figure}

As usual, the phase diagrams are computed by minimizing Eq.~(\ref{effective_potential_finiteT_SMC_side_3d}) with respect to $\sigma$ and $\eta$. In all numerical calculations in $(2+1)d$ case,  we have fixed $g_e = 1.1 > g_\Lambda = 1.0 > g_c = 0.9$.  We have also taken $M=1.0$ in all cases,  to be consistent with the (1+1)d case discussed previously.

In Fig.~\ref{fig:8}~(a) we show the behavior of $\sigma$ and $\eta$ as a function of $m$ at zero temperature. This behavior is very similar to the (1+1)d case for both order parameters.
\begin{figure}[b]\centering
  \includegraphics[width=\columnwidth]{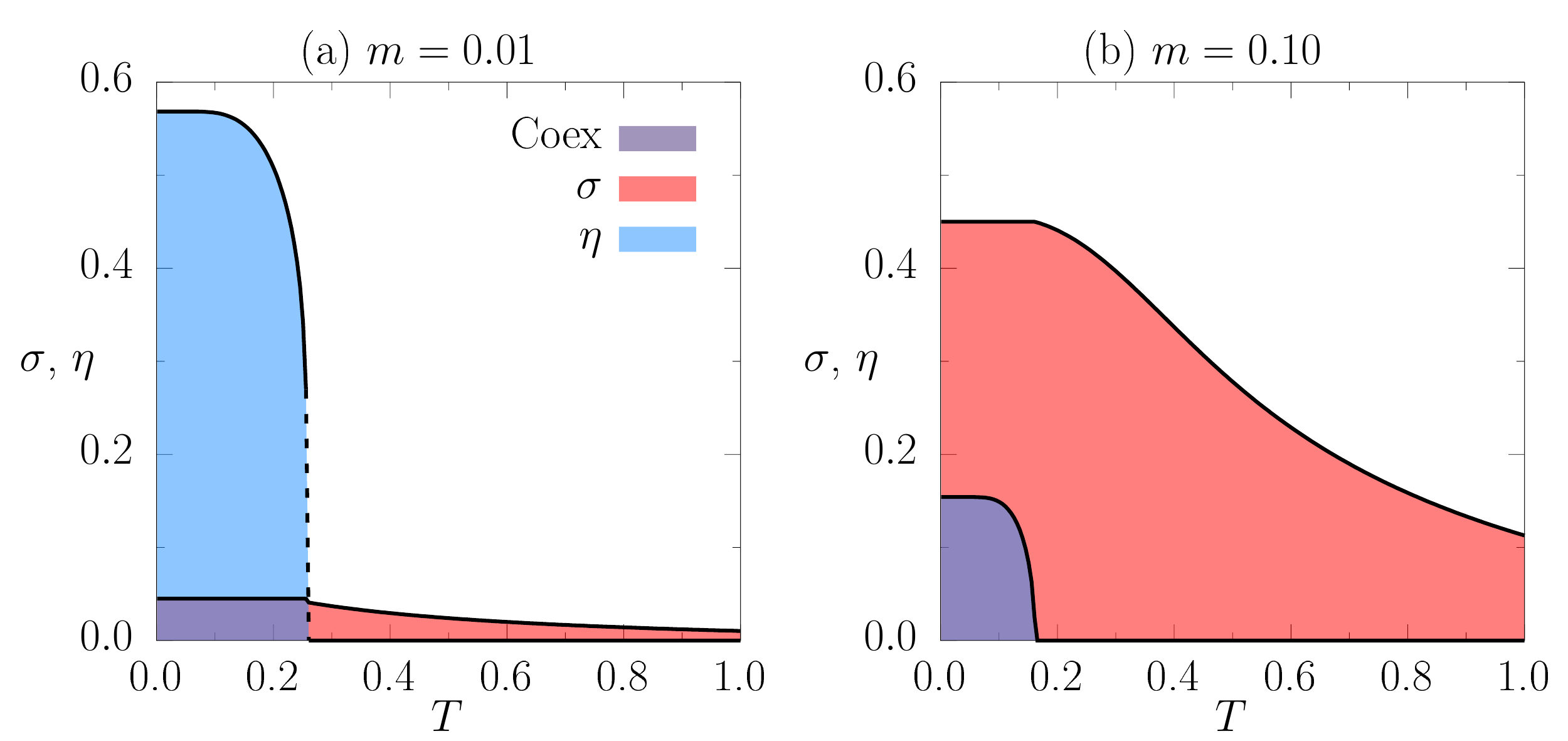}
  \caption{\label{fig:9} (Color online) Order parameters as a function of $T$ for small values of mass (bare \textit{gap}). (a) For $m = 0.01$, the system undergoes a first-order (dashed line) finite temperature phase transition, which is very different when compared with its (1+1)d version, see Fig.~\ref{fig:2}~(a). (b) For $m = 0.10$, i.e., increasing mass (\textit{gap}), but still in the region where $\eta \neq 0$ the finite temperature transition becomes second-order (continuous line). Again, the value of $\sigma$ (\textit{gap}) remains constant while $\eta \neq 0$, consistent with the emergence of an additional \textit{gap} in the EI phase.} 
\end{figure}

In Figs.~\ref{fig:9}~(a) and~(b) we show the order parameters $\sigma$ and $\eta$ as a function of $T$, for small fixed values of $m$.   It is interesting to note that for small values of $m$, Fig.~\ref{fig:9}~(a), we have a first-order (dashed line) thermal transition for $\eta$, which is different from the $(1+1)d$ case. For large values of $m$,  in the region of $\eta \neq 0$, Fig.~\ref{fig:9}~(b), we recover the second-order character of the transition for $\eta$. However, we also have $\sigma$ constant while $\eta \neq 0$, similarly to the (1+1)d case, which is, again, consistent with the same discussion of the emergence of an additional \textit{gap} in the EI phase.

\begin{figure}[t]\centering
  \includegraphics[width=0.85\columnwidth]{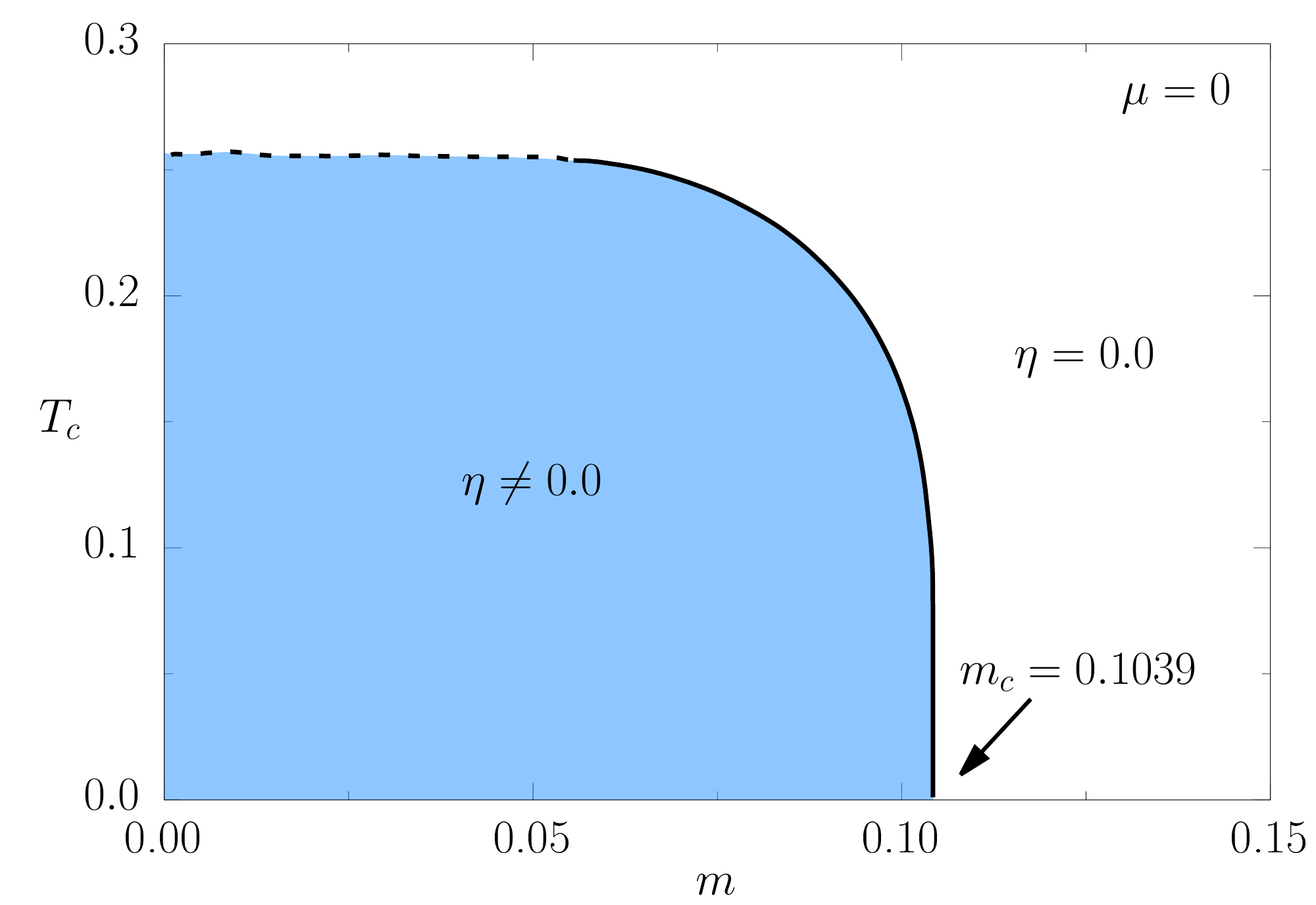}
  \caption{\label{fig:10} (Color online) Critical temperature for $\eta$ as a function of mass at the the SMC regime for (2+1)d case. Note that now there is a region of small mass where the systems undergoes a first-order (dashed line) thermal transition. For large values of $m$ wherein $\eta \neq 0$, the finite temperature phase transition becomes second-order (continuous line). Also there is a critical mass ($m_c$) at zero temperature, where the EI phase disappears and this value is independent of $\mu$, see caption of Fig.~\ref{fig:4}.} 
\end{figure}

The critical temperature for $\eta$ as a function of $m$ (bare \textit{gap}) in the SMC regime is shown in Fig.~\ref{fig:10}. One can see that there is a region of small mass (bare \textit{gap}) where the system  undergoes a first-order transition (dashed line).  There is also  a critical mass ($m_c$) at zero temperature where the EI phase disappears. In other words,  for a fixed $g_e$ we can not increase the bare and the renormalized \textit{gaps} of the system indefinitely. Otherwise, $\eta \rightarrow 0$. The first-order character of the EI critical temperature for small masses is directly associated with dimensional effects,  since at zero temperature we obtain very similar results for the EI phase in 2d as well as 3d cases, see Figs.~\ref{fig:1} and~\ref{fig:8}.

Following the same lines of the previous sections, we describe the SM regime by linearizing the dispersion relation at the crossing  point.   Thus,  we consider, again, two massless generalized GN models in which $k_0$ is a measure of the band \textit{overlaps}. Applying the Large $N$ limit approach to the GN model in (2+1)d for the SM regime of the EI phase, we obtain,
\begin{eqnarray}
V_{eff}^{N}&=&\frac{\left(\sigma^2+\eta^2\right)^{3/2}}{6 \pi}+\left(\frac{1}{g_c}-\frac{1}{g_\Lambda}\right)\frac{\sigma^2}{2} + \left(\frac{1}{g_e}-\frac{1}{g_\Lambda}\right)\frac{\eta^2}{2} \nonumber
\\
&+&\frac{k_0}{2\pi^2}(\sigma^2+\eta^2)\left[\ln\left(\frac{\sigma^2+\eta^2}{M^2}\right)-3\right]\nonumber
\\
&-&\frac{T}{\pi}\int_{0}^{\infty}dx \ x\left\{\ln\left(1+e^{-\frac{E-\mu}{T}}\right)+\mu \rightarrow -\mu\right\}
\label{effective_potential_finiteT_SM_side_3d}
\end{eqnarray}
where $k_0 < 0$, $g_\Lambda = 3\pi^2/(2\Lambda)$ and $E^2 = x^2+\sigma^2+\eta^2$. 

\begin{figure}[t]\centering
  \includegraphics[width=0.85\columnwidth]{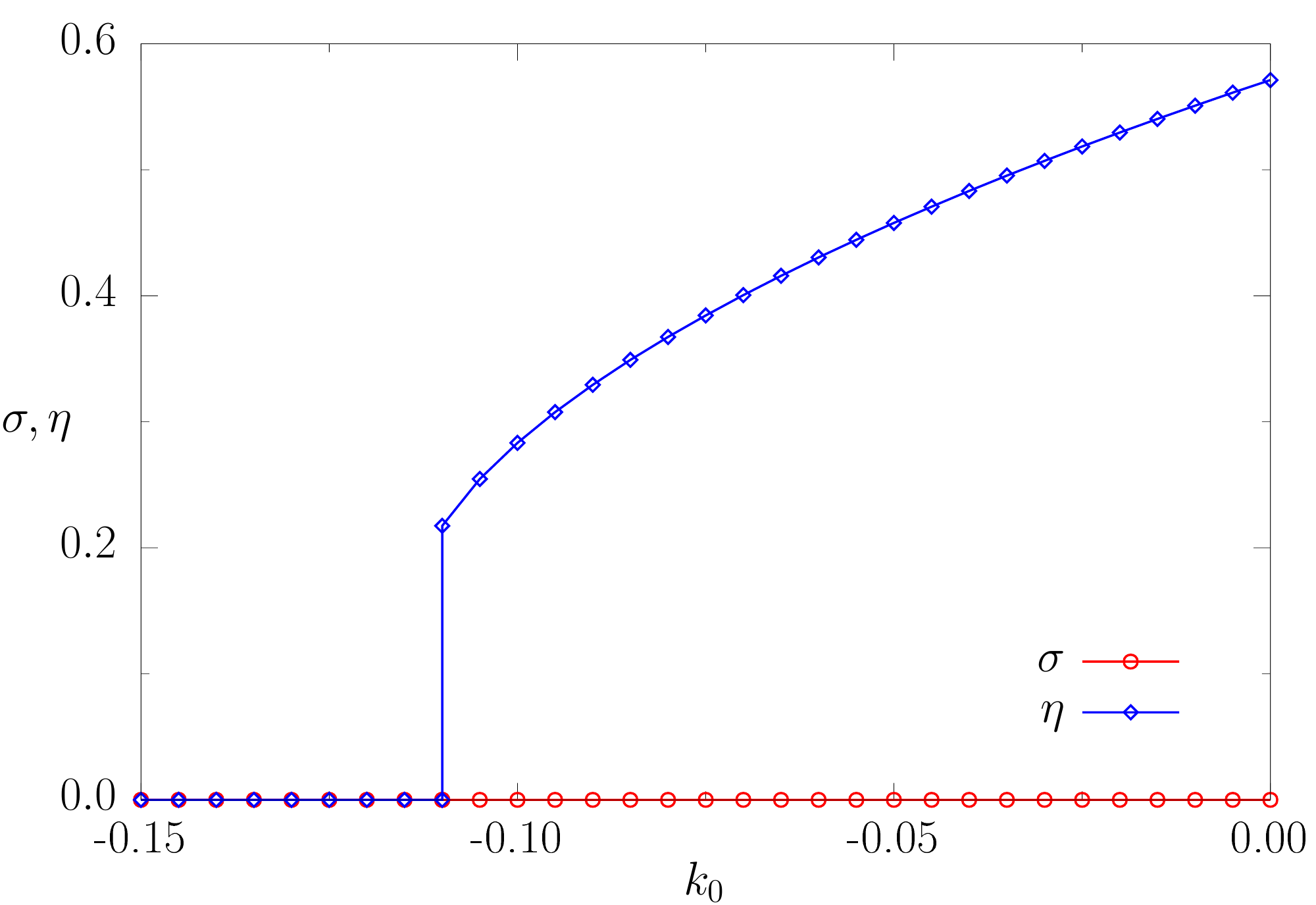}
  \caption{\label{fig:11} (Color online) Order parameters as a function of $k_0$ (overlap), from the minimization of Eq.~(\ref{effective_potential_finiteT_SM_side_3d}), at zero temperature. Note that $\sigma =0$ (red circles), while $\eta$ (blue squares) undergoes a quantum first-order (blue square line) transition as we increase $\left|k_0\right|$.} 
\end{figure}

It is worth to emphasize that we only renormalize the divergence coming from $k_0$, since $g_\Lambda$ can be seen as a natural \textit{cutoff} of the system, related, for instance, to the lattice parameter. By minimizing  Eq.~(\ref{effective_potential_finiteT_SM_side_3d}) at zero temperature,  it is simple to confirm that  $\sigma = 0$ for all the  SM regime,  while $\eta$ is finite for small $k_0 \leq 0$ and goes to zero abruptly as we increase $\left|k_0\right|$. This is a clear  signature that the system undergoes a quantum first-order (dashed line) transition,  see Fig.~\ref{fig:11}. 

\begin{figure}[b]\centering
  \includegraphics[width=0.85\columnwidth]{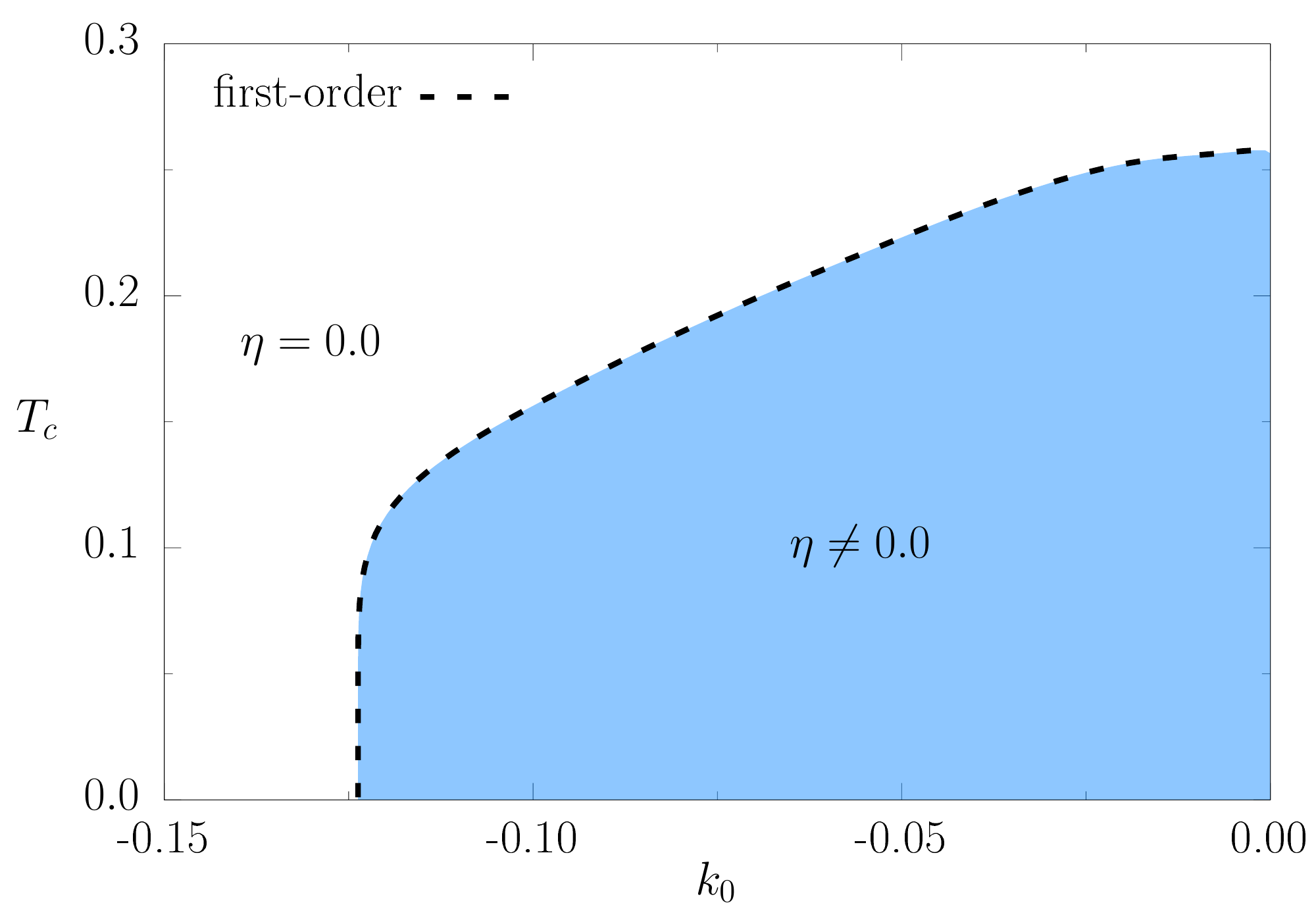}
  \caption{\label{fig:12} (Color online) Critical temperature for $\eta$ as a function of $k_0$. All the region where $\eta \neq 0$ (blue) exhibits a thermal first-order (dashed line) transition. Note that we have a small region of $k_0 \leq 0$ for the emergence of the EI that is very similar when compared to the region in which $\eta \neq 0$ is more stable in (1+1)d, see Fig.~\ref{fig:6}.} 
\end{figure}

In Fig.~\ref{fig:12}, including effects of finite temperature, analogously to the (1+1)d case, we obtain that there is a small region of $k_0 \leq 0$ where the EI may appear, i.e., $\eta \neq 0$. Our numerical results show that all this region is a region where the system exhibits a thermal first-order (dashed line) transition, very similar to the region where $\eta \neq 0$ is more stable in Fig.~\ref{fig:6}.
\begin{figure}[t]\centering
  \includegraphics[width=0.85\columnwidth]{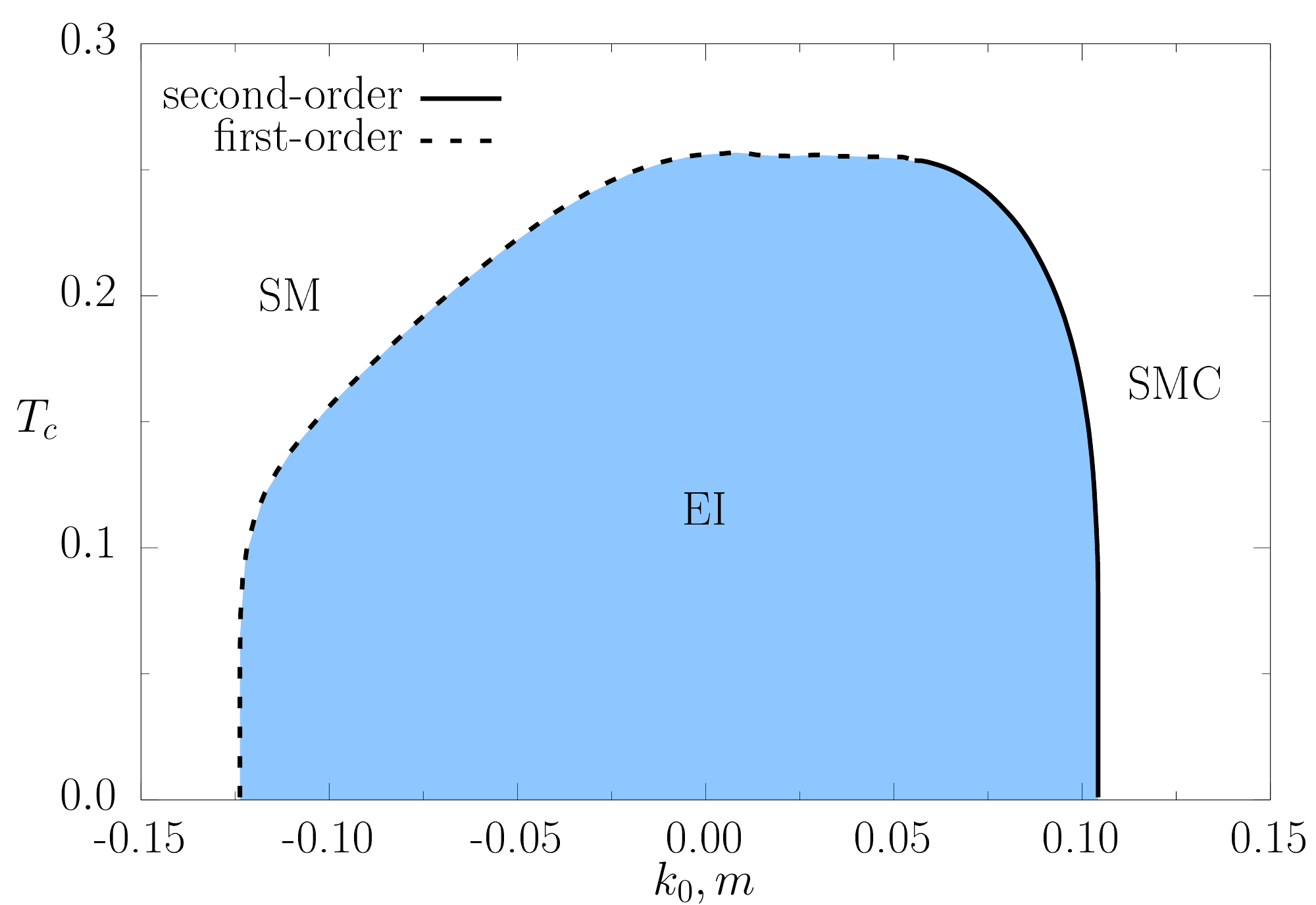}
  \caption{\label{fig:13} (Color online) EI phase diagram combining the two models in (2+1)d. Again, In the x-axis we have $k_0$ (negative) for the SM regime, which is associated with the overlap between bands, and $m$ (positive) for the SMC regime, that is related to the bare \textit{gap} of the system. Continuous line denotes second-order phase transition, while dashed line denotes first-order ones.} 
\end{figure}
\begin{figure}[b]\centering
  \includegraphics[width=0.85\columnwidth]{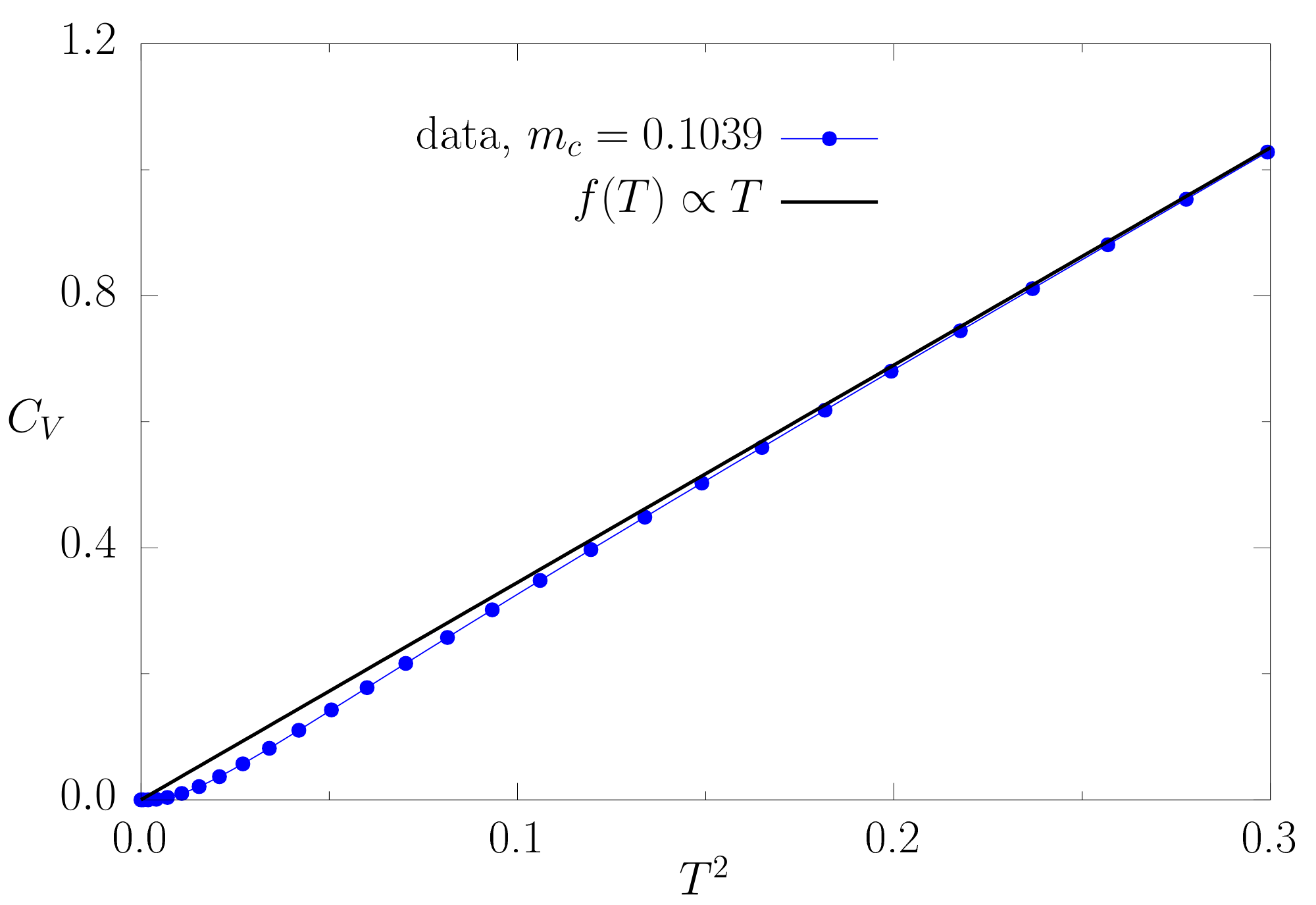}
  \caption{\label{fig:specific_heat_3d} (Color online) Specific heat, at constant volume, as a function of $T^2$ above the EQCP, i.e., at the fine-tuned value of $m_c=0.1039$. One can see that for large $T$, $C_V$ exhibits a linear behavior, consisted with the expected behavior at a QCP, $C_V/T \propto T^{(d-z)/z}$ with $d=2$ and $z=1$. However, at low temperatures the specific heat is thermally activated due to the presence of the \textit{gaps} ($\sigma \neq 0$ and $m \neq 0$), deviating from the linear behavior.} 
\end{figure}

Finally, combining Fig.~\ref{fig:10} and Fig.~\ref{fig:12}, we  obtain the complete  EI phase diagram from our models in (2+1)d case.  We show this result in Fig.~\ref{fig:13}. Note that, again, we have a step-like shape for the EI phase diagram, although we no longer have the minimum of the effective potential always at the origin in the EI phase for larges values of $\left|k_0\right|$. In other words, for the (2+1)d case, the EI state might appear only for small \textit{overlap} and small \textit{gaps} of the system, in agreement with the expected behavior of the EI phase~\cite{Mott, Knox}. The main difference within our model, which neglects curvature effects in 2d as well as 3d, is the character of first-order thermal phase transition depending on whether the EI is approached from a SM or a SMC phase, see Fig.~\ref{fig:7} and Fig.~\ref{fig:13}. Also observe that we find, again, an EQCP for (2+1)d at the SMC regime.

Therefore, in Fig.~\ref{fig:specific_heat_3d},  we show the specific heat, at constant volume, as a function of $T^2$ for the fined-tuned value of mass $m=m_c$. For large $T$, now we obtain a quadratic behavior for the specific heat at constant volume, which is consistent with the scaling prediction, $C/T \propto T^{(d-z)/z}$ in $d=2$ with $z=1$. On the other hand, for the low $T$ regime, we also obtain a dominant exponential thermally activated term besides the power law contributions due to quantum critical effects, which deviates from the scaling prediction at large $T$, as shown in Fig.~\ref{fig:specific_heat_3d}  (see fitting curve $f(T)$). Again, this is a direct consequence of the unusual character of the EQCP.

We emphasize that this deviation of $C_V$ from the quadratic scaling behavior at low $T$, is more subtle when compared to the (1+1)d case, see Fig.~\ref{fig:specific_heat_2d}. Note that we need to further cool down the temperature of the system to observe this deviation due to the additional exponential thermally activated term on the specific heat. We attribute this behavior to dimensional effects on quantum systems, since in low dimensional systems fluctuations become pronounced.

\section{Conclusions}
\label{sec:conclusions}

The excitonic state is an elusive state of matter. It involves condensation of chargeless particles with small impact on the transport properties of the material. As it presents a strong theoretical possibility, it has been intensively sought in nature. Recently,  a strong candidate for an EI has been found, namely the system Ta$_2$NiSe$_5$~\cite{nature2,Wakisaka,Volkov,Sugimoto,disalvo,Kim}.

In this paper we use an extended version of an exactly soluble model of QFT to describe an EI in one and two-spatial dimensions. The origin of this state lies in  the electronic correlations between quasi-particles in different bands of the model. We have identified an inter-band order parameter that characterizes the excitonic state. It is different from that associated with  the breakdown of  chiral symmetry that usually occurs in GN models. The nature of the order parameter in the excitonic state implies that charge is not conserved in individual bands but only globally in the two-band system. It is related to the appearance of a spontaneous hybridization in the system. Notice that hybridization involves the \textit{overlap} of different orbitals and consequently it can  change due to  variation  in  atomic positions as, for example, by applying pressure in the system, or due to structural changes that modify  this mixing. A bare hybridization acts as a conjugate field to the EOP and may destroy this transition.

We considered two different situations where the excitonic state may arise; in a SM with small band \textit{overlap} and  in a SMC with a small \textit{gap} between the valence and conduction bands. In the former case we obtain that the instability to the excitonic state is a discontinuous transition, although a very weak one. We have introduced a parameter that characterizes the \textit{overlap} between the bands in the SM regime. It is related to the density of carriers in the system and when it is small, it implies a poor screening of the charges, such that the electron-hole attraction is effective. For large values of this parameter the charges are sufficiently screened and the excitonic state is destroyed.

In the SMC region we obtain that  there is a minimum value of the inter-band interaction to produce the excitonic state, as expected. In this case the transition may be second-order and at zero temperature it gives rise to a special kind of quantum critical point. Since the main role of the interactions is to renormalize the \textit{gap} between bands, the system remains gapped even at the EQCP. The critical power law corrections to the thermodynamic properties at quantum criticality appear on top of an exponentially activated contribution due to the presence of the bare and/or the renormalized \textit{gaps}.

Our approach provides a description of the excitonic transition in qualitative agreement with the experimental observations in the system Ta$_2$NiSe$_5$~\cite{nature2,Wakisaka,Volkov,Sugimoto,Kim}. In this system the excitonic instability that occurs with decreasing temperature is accompanied by the appearance of a renormalized \textit{gap} and a flattening of the top of the valence band. Both features are predicted in our model for the SMC-EI transition.

\section{Acknowledgments}

We would like to thank the Brazilian agencies \textit{Funda\c c\~ao Carlos Chagas Filho de Amparo \`a  Pesquisa do Estado do Rio de Janeiro} (FAPERJ), \textit{Coordena\c c\~ao de Aperfei\c coamento de Pessoal de N\'\i vel Superior} (CAPES) - Finance Code 001 and \textit{Conselho Nacional de Desenvolvimento Cient\'\i fico e Tecnol\' ogico} (CNPq) for partial financial support. N.L. would like to thank the FAPERJ for the post doctoral fellowship of the \textit{Programa de P\' os-Doutorado Nota 10 - 2020} (E-26/202.184/2020) as well as for the \textit{Bolsa de Bancada para Projetos} (E-26/202.185/2020).



\end{document}